\documentclass[12pt,a4paper]{article}
\pdfoutput=1
\usepackage{jheppub}
\usepackage{epstopdf}
\usepackage{graphicx}
\usepackage{epsfig}
\usepackage{dcolumn}  
\usepackage{bm}    
\usepackage{amssymb} 
\usepackage{amsmath,bm}
\usepackage{amsfonts}    
\usepackage{slashed}  
\usepackage[mathscr]{euscript}
\usepackage{epsfig}
\usepackage[position=t,singlelinecheck=off]{subfig}
\usepackage{caption}
\title{Higher spin entanglement entropy from CFT}

\author[a]{Shouvik Datta,} 
\author[a]{Justin R.~David,} 
\author[b]{Michael Ferlaino} 
\author[a,b]{and S.~Prem Kumar} 
\affiliation[a]{Centre for High Energy Physics, Indian Institute of Science,\\ C. V. Raman Avenue, Bangalore 560012, India.}
\affiliation[b]{Department of Physics, Swansea University,\\Singleton Park, Swansea SA2 8PP, UK.}

\emailAdd{shouvik, justin@cts.iisc.ernet.in}
\emailAdd{\\\qquad\quad\, pymf, s.p.kumar@swansea.ac.uk}
\abstract{We consider free fermion and free boson CFTs in two dimensions, deformed by a chemical potential $\mu$ for the spin-three current. For the CFT on the infinite spatial line, we calculate the finite temperature entanglement entropy of a single interval perturbatively to second order in $\mu$ in each of the theories. We find that the result in each case is given by the same non-trivial function of temperature and interval length. Remarkably, we further obtain the same  formula using a recent Wilson line proposal for the holographic entanglement entropy, in holomorphically factorized form, associated to the spin-three black hole in ${\rm SL}(3,{\mathbb R}) \times {\rm SL}(3,{\mathbb R})$ Chern-Simons theory. Our result suggests that the order $\mu^2$ correction to the entanglement entropy may be universal for ${\cal W}$-algebra CFTs with spin-three chemical potential, and constitutes a check of the holographic entanglement entropy proposal for higher spin theories of gravity in AdS$_3$.
}

\begin{document}
 \maketitle 

\def\be{\begin{equation}}
\def\ee{\end{equation}}
\def\bea{\begin{eqnarray}}
\def\eea{\end{eqnarray}}
\def\nn{\nonumber}
\def\pd{\partial}
\def\Re{R\'{e}nyi }
\def\l1{{\text{1-loop}}}
\def\uy{u_y}
\def\ur{u_R}
\def\o{\mathcal{O}}
\def\Cl{{{cl}}}
\def\bz{{\bar{z}}}
\def\by{{\bar{y}}}
\def\bX{\bar{X}}
\def\im{{\text{Im}}}
\def\re{{\text{Re}}}
\def\cn{{\text{cn}}}
\def\sn{{\text{sn}}}
\def\dn{{\text{dn}}}
\def\K{\mathbf{K}}
\def\n1{\Bigg|_{n=1}}
\def\fin{{\text{finite}}}
\def\R{{\mathscr{R}}}
\def\one{{(1)}}
\def\zero{{(0)}}
\def\n{{(n)}}
\def\tr{\text{Tr}}
\def\T{\mathcal{T}}
\def\TT{\tilde{\mathcal{T}}}
\def\O{\mathcal{O}}
\def\cN{\mathcal{N}}
\def\P{\Phi}
\def\csch{{\text{cosech}}}
\def\W{{\tilde{W}}}
\def\T{{\tilde{T}}}
\def\by{\bar{y}}
\newcommand*\xbar[1]{%
  \hbox{%
    \vbox{%
      \hrule height 0.5pt % The actual bar
      \kern0.5ex%         % Distance between bar and symbol
      \hbox{%
        \kern-0.1em%      % Shortening on the left side
        \ensuremath{#1}%
        \kern-0.1em%      % Shortening on the right side
      }%
    }%
  }%
}

\section{Introduction and Summary}
Entanglement is a fundamental property of quantum systems.  
Entanglement entropy (EE) provides a powerful measure of this fundamental property, particularly for extended systems with many degrees of freedom close to a critical point \cite{Calabrese:2004eu, Calabrese:2005in}\footnote{For earlier works on entanglement entropies in 1+1 QFTs, see 
\cite{Korepin:2004zz}}. It acts as a measure of the effective degrees of freedom within a given region interacting with the rest of the theory. The AdS/CFT correspondence \cite{maldacena, witten} has led to an exciting  avenue for computing EE in strongly coupled field theories using the elegant geometric prescription due to Ryu and Takayanagi \cite{Ryu:2006bv, Ryu:2006ef}. Recently,  remarkable progress has been made in the development of holographic descriptions of field theories with higher spin conserved currents \cite{Klebanov:2002ja, Giombi:2009wh, Giombi:2010vg,gg}. Specifically, the proposal due to Gaberdiel and Gopakumar relates ${\cal W}_N$ minimal model CFTs in two dimensions in a 't Hooft-like large-$N$ limit, to classical theories of higher spin gravity in AdS$_3$ \cite{gg, Gaberdiel:2012ku, Gaberdiel:2012uj}. This presents a  tractable framework for exploring and understanding holography within the novel setting of higher spin gravity theories, in particular the hs[$\lambda$] theory \cite{Prokushkin:1998bq} in AdS$_3$. This holographic duality has raised fascinating questions regarding the nature of holographic thermodynamics and black hole-like objects in higher spin gravity \cite{Gutperle:2011kf, Ammon:2011nk}, and whether the notion of the holographic 
EE can be appropriately generalized to such theories 
\cite{deBoer:2013vca, Ammon:2013hba}.

The aim of this paper is to consider the simplest CFTs in two dimensions which have a ${\cal W}_\infty[\lambda]$ symmetry, namely the free boson $(\lambda=1)$ and free fermion $(\lambda=0)$ theories \cite{Pope:1989ew, Bergshoeff:1990yd, Pope:1991ig, Bakas:1990ry}, and to compute the finite temperature entanglement entropy in these theories in the presence of a chemical potential for higher spin charge.

 Our primary motivation is to eventually make contact with the holographic description of such charged, thermal states in CFTs with  
${\cal W}$-algebra symmetries. In particular these should correspond, in the limit of large central charge, to black hole solutions in higher spin theories of gravity on AdS$_3$, carrying higher spin charges. The simplest such examples to which most attention has been devoted are the spin-three black holes in ${\rm SL}(3,{\mathbb R})\times {\rm SL}(3,{\mathbb R})$ Chern-Simons theory \cite{Gutperle:2011kf, Ammon:2011nk} and within hs[$\lambda$] theory \cite{Kraus:2011ds}. The asymptotic symmetries of the gravity backgrounds for these examples are, respectively, the ${\cal W}_3$ and ${\cal W}_\infty[\lambda]$ algebras \cite{Henneaux:2010xg, campoleoni}. The former is generated by the stress tensor and a spin-three current, whilst the latter is generated by an infinite set of higher spin currents with spin $s\geq 2$.

Whilst many aspects of the thermodynamics of higher spin black holes have been understood \cite{Gutperle:2011kf,Perez:2012cf, Campoleoni:2012hp, Perez:2013xi, deBoer:2013gz} and impressively matched with CFT computations \cite{Kraus:2011ds, Gaberdiel:2012yb, Gaberdiel:2013jca}, much still remains to be understood. Proposals for the definition of a holographic entanglement entropy in higher spin theories dual to CFTs with ${\cal W}$-symmetry have been put forth independently by two different works \cite{deBoer:2013vca, Ammon:2013hba}. 
Given that higher spin theories in AdS$_3$ can be formulated in the language of Chern-Simons theory,
both proposals naturally revolve around the computation of Wilson lines in certain representations. Using their respective prescriptions the works of \cite{deBoer:2013vca, Ammon:2013hba} have been able to calculate the entanglement entropy for spin-three black holes in ${\rm SL}(3,{\mathbb R})\times {\rm SL}(3,{\mathbb R})$ Chern-Simons theory, associated to a boundary ${\cal W}_3$ CFT. However, there are no existing corresponding results from CFT to compare 
with\footnote{Two interesting papers \cite{Chen:2013dxa,Perlmutter:2013paa}
have recently appeared addressing one-loop corrections to 
R\'enyi and entanglement entropies in higher spin theories, 
in the absence of chemical potential for higher spin charges. 
}.In particular, one needs to compute the finite temperature entanglement entropy (of one or more intervals) in a CFT with ${\cal W}$-symmetry, in the presence of a chemical potential for spin-three charge. This is precisely the calculation we set out to address in this paper within the free fermion and free boson CFTs which are tractable examples with ${\cal W}_\infty$ symmetry.

The procedure for calculating the entanglement entropy of free boson and free fermion CFTs in two dimensions is well understood {\em at the conformal point} \cite{Calabrese:2004eu, Calabrese:2005in, Calabrese:2009qy, Casini:2009sr}. However, the calculation that we want to perform involves a deformation of the CFT by a chemical potential $\mu$ for the spin-three current, which we will treat as a perturbation. The chemical potential is a coupling with dimension $-1$ and therefore the dimensionless small parameter is actually $\mu T\ll 1$, where $T=\beta^{-1}$ is the temperature. We will therefore perform conformal perturbation theory to obtain the corrections to the entanglement entropy at order $\mu^2$.

An important feature of the conformal perturbation theory that we study is that it is ``holomorphically factorized'' in the sense that the perturbing operator is a sum of the $(3,0)$ holomorphic spin-three current and its anti-holomorphic counterpart. From the dual gravity side this is a somewhat unnatural way of introducing a chemical potential and leads to the so-called holomorphic formulation of thermodynamics \cite{Gutperle:2011kf, Ammon:2011nk, deBoer:2013gz} as opposed to the canonical one \cite{Perez:2012cf, Campoleoni:2012hp, Perez:2013xi,Compere:2013nba}. It is however understood that the two formulations are related \cite{deBoer:2013gz}. 

The holographic higher spin EE proposal of \cite{deBoer:2013vca} in particular also includes a proposal for the holomorphically factorized picture which we will refer to as $S_{\rm EE}^{\rm holo}$. Their proposed formula, based on computing a Wilson line (see eqs.\eqref{hol-dB-J}, \eqref{funct-W}) has only been evaluated for the spin-three black hole solution of Gutperle and Kraus \cite{Gutperle:2011kf} within ${\rm SL}(3,{\mathbb R})\times {\rm SL}(3,{\mathbb R})$ Chern-Simons theory.

Before proceeding further, we can state the main result of our work. We first express the entanglement entropies as a power series in $\mu^2$, so that $S_{\rm EE}\,= S^{(0)}_{\rm EE}\,+\,\mu^2\,S^{(2)}_{\rm EE}\,+\ldots$. 
Calculating the entanglement entropy $S_{\rm EE}$ of a single interval of length $\Delta$ within the free boson and free fermion theories to ${\cal O}(\mu^2)$, at finite temperature and on the infinite spatial line, we find 
\bea
&&\left(S^{(0)}_{\rm EE} +\mu^2S^{(2)}_{\rm EE}\right)\Big|_{\rm boson}\,=\,\left(S^{(0)}_{\rm EE} +\mu^2 S^{(2)}_{\rm EE}\right)\Big|_{\rm fermion}\,=\, \left(S^{(0)}_{\rm EE} +\mu^2S^{(2)}_{\rm EE}\right)\Big|_{\rm sl(3) \oplus sl(3)}^{\rm holo}\label{EEtotal}\nonumber\\\nonumber\\
&&=\,\frac{c}{3}\log\left|\frac{\pi}{\beta}\sinh\left(\frac{\pi\Delta}{\beta}\right)\right|\,+\,c\,\frac{\mu^2}{\beta^2}
\left[\frac{32\pi^2}{9}\,\left(\tfrac{\pi\Delta}{\beta}\right)\,\coth\left(\tfrac{\pi\Delta}{\beta}\right)\,-
\,\frac{20\pi^2}{9}\right.\\\nonumber\\\nonumber
&&\qquad\qquad\qquad\left.-\,\frac{4\pi^2}{3}\csch^2\left(\tfrac{\pi\Delta}{\beta}\right)\,
\left\{\left(\tfrac{\pi\Delta}{\beta}\coth\left(\tfrac{\pi\Delta}{\beta}\right)\,-\,1\right)^2\,+\,\left(\tfrac{\pi\Delta}{\beta}\right)^2\right\}\right]\,.
\eea
First of all, this is an unexpected agreement across distinct (perturbed) CFTs that cannot persist beyond quadratic order in $\mu$ simply because at high temperatures the entanglement entropy must become extensive in the interval length $\Delta$ and reproduce the thermal entropy of the respective deformed CFTs. It is known that beyond second order in $\mu$ the thermal entropy for ${\cal W}_\infty[\lambda]$ theories 
depends non-trivially on $\lambda$ \cite{Kraus:2011ds, Gaberdiel:2012yb}, whilst the order $\mu^2$ correction to thermal entropy is independent of $\lambda$.

The requirement that \eqref{EEtotal} should reproduce the thermal entropy correction at high temperatures, $\Delta/\beta\gg 1$, is guaranteed by the first term in the order $\mu^2$ correction in eq.\eqref{EEtotal}. The remaining contributions are suppressed (exponentially) at high temperature and it is {\em a priori} not necessary that these pieces should agree across theories with different ${\cal W}$-symmetries. We further note that the field theory calculation we perform is at fixed central charge for two theories with ${\cal W}_\infty[0]$ and ${\cal W}_\infty[1]$ symmetry\footnote{The free fermion theory  actually has  ${\cal W}_{1+\infty}$ symmetry due to the presence of a U(1) current. We will discuss this in more detail in subsequent sections.}
 while the holographic result is for a theory with ${\cal W}_3$ symmetry and central charge $c\to \infty$. For these reasons the appearance of an identical formula for these three different theories is suggestive of a universal result, although we do not have any additional evidence to suggest that it is so.

The basic idea behind the field theory calculation is to employ the replica trick to compute the  R\'enyi entropies and in the end take a limit wherein the number of replicas approaches unity to finally obtain the entanglement entropy. It is worth noting that the R\'enyi entropies for the free boson and free fermion theories do not match. However, in the limit which yields the entanglement entropy, the two theories exhibit identical results. The central element of the calculation relies on establishing three- and four-point functions involving single and double insertions of spin-three currents, respectively, in the presence of twist and anti-twist operators. It is the twist operators that enable the computation of the replica partition function on a multi-sheeted Reimann surface. These elements of the calculation are quite different for the free fermion and free boson theories. In fact, while the twist fields are explicitly known for free fermions, this is not the case for the bosons where the computation is much more involved, and yet the two theories yield the same final result.

The outline of the paper is as follows: In section 
\ref{holpert} we briefly review known aspects of higher spin black hole solutions and explain how to obtain the first thermal corrections to the partition function using conformal perturbation theory. In section \ref{pertEE} we present the basic ingredients required for the perturbative evaluation of the entanglement entropy. We also present the result of our calculations and list its features. Section \ref{freefer} is devoted to the computation of the R\'enyi entropies for the free fermion theory on the cylinder. The free boson computation including the detailed derivation of the requisite three- and four-point functions is presented in section \ref{freebose}. The holographic proposal for EE is reviewed in section \ref{holEE} and the result for the ${\rm SL}(3)$ Chern-Simons theory is written down explicitly. In section \ref{cosetting} we explore the possibility of removing the extra U(1) current in the free fermion theories by a cosetting 
procedure and find the appearance of ${\cal W}_N$-algebras. We further explore the calculation of EE in the free field realization of ${\cal W}_N$-algebras. In appendix \ref{walgebra} we list some basic ${\cal W}$-algebra OPEs, while appendix \ref{integrals} is devoted to the detailed evaluation of the integrals we encounter in our perturbative calculations.

%----------------------------

%\begin{figure}[!t]
%\begin{tabular}{c}
%\includegraphics[width=6in]{lambda.pdf}
%\end{tabular}
%\caption{\small \textbf{The space of $\mathcal{W}_{\infty} [\lambda]$
%theories.} The entanglement entropy is calculated from CFT at the points
%$\lambda=0$ and 1 which corresponds to the free fermion and the free boson.
%At the point $\lambda=-3$ we calculate the entanglement entropy
%holographically using the Wilson-line prescription. The
%$\mathcal{O}(\mu^2)$ correction to the EE at all these three points exactly
%match providing a check of the holographic proposal and also furnishing a
%strong evidence for universality at this order.}
%\label{fig1}
%\end{figure}

%-----------------------------
\section{Higher spin chemical potentials and thermodynamics}
\label{holpert}
Black hole solutions in higher spin theories of gravity in ${\rm AdS}_3$, carrying higher spin charges, were first constructed by Gutperle and Kraus in \cite{Gutperle:2011kf}. The Gutperle-Kraus solutions were obtained within ${\rm SL}(3,{\mathbb R}) \times {\rm SL}(3,{\mathbb R})$  Chern-Simons theory which is a higher spin theory with only a spin-3 field in addition to gravity. The solutions obtained within this framework could be interpreted as deformations of the boundary CFT$_2$ by a chemical potential for the spin-3 current. 

The chemical potential deformation in the boundary CFT has two important features 
\cite{Gutperle:2011kf, Ammon:2011nk} :
The spin-three current being  a dimension three operator, a chemical potential for the corresponding charge appears as an irrelevant coupling. Secondly, comparison of the  Ward identities in the deformed CFT and the Chern-Simons equations of motions in the bulk indicates that the deforming operator is the holomorphic $(3,0)$ current (plus its anti-holomorphic counterpart).

This interpretation of the higher spin deformation leads to the so-called `holomorphic' formulation of black hole thermodynamics which has so far provided the most natural route for successful comparison of bulk gravity observables with  CFT  results\cite{Kraus:2011ds, Gaberdiel:2012yb, Gaberdiel:2013jca}.  There also exists a different, `canonical' formulation of bulk thermodynamics which is more natural from the viewpoint of gravity and holography \cite{Perez:2012cf, Campoleoni:2012hp, Perez:2013xi,Compere:2013nba,deBoer:2013gz}. It has however been pointed out in \cite{deBoer:2013gz} that the two formulations are related. 

In this paper we will exclusively discuss the holomorphic approach. It would be very interesting to revisit the results in this paper in light of the  recent proposal to introduce higher spin chemical potentials via the temporal component of the Chern-Simons connections \cite{Henneaux:2013dra}.

In the formulation of \cite{Gutperle:2011kf}, a chemical potential $\mu$ for higher spin charge deforms the CFT$_2$ by a dimension three operator
\be
\delta I \,=\, -\mu\,\int d^2z\,\left(W(z)\,+\, \overline{{ W}}(\bar z)\right)\,,\label{action}
\end{equation}
where $W(z)$ is the holomorphic spin-3 current on the plane. We expect that for small $\mu$ (relative to the inverse energy scale or the inverse temperature), all observables should be computable via conformal perturbation theory in $\mu$. 

The perturbative (in $\mu$) approach has been applied in the language of canonical quantization, in conjunction with modular invariance of the torus partition function, to compute CFT$_2$ thermodynamics \cite{Gaberdiel:2012yb} and to demonstrate non-trivial agreement with the higher spin black holes of \cite{Gutperle:2011kf}. However, the same excercise has not been carried out directly in the Lagrangian/path-integral language of perturbation theory. 
In this paper we will directly  compute the finite temperature R\'enyi and entanglement entropies for certain free CFT's. To this end we begin by outlining our approach and our conventions by first calculating the order $\mu^2$ correction to the high temperature free energy of the perturbed CFT.

\subsection{`Holomorphic' perturbation theory and free energy at ${\cal O}(\mu^2)$} 
To order $\mu^2$, the deformed CFT partition function is given in conformal perturbation theory as
\bea
{Z}\,=&&\,{Z}^{(0)}_{\rm CFT}\times\\\nonumber
&&\left(1\,-\,\mu\int d^2 z\,\langle W \rangle_{\rm CFT}\,+\,
\tfrac{1}{2}\mu^2 \int d^2z_1\int d^2z_2\, \langle W(z_1)\, W(z_2)\rangle_{\rm CFT} \,+\ldots {\rm h.c.}\right)
\eea
It is worth noting that conformal perturbation theory in a purely holomorphic (or anti-holomorphic) operator is quite unusual and non-standard\footnote{We find that, at least to the order we have worked in, perturbation theory in the holomorphic operator leads to finite results, requiring no UV regularization prescription. Working on the cylinder does appear to provide an IR regularization prescription because the same perturbation theory on the infinite plane seems to be sensitive to IR boundary conditions.}.

We will always take the CFT to be on the infinite spatial line, and view this as the high temperature limit of the CFT on a spatial circle. At any finite temperature $T= 1/\beta$, 
our strategy will be to infer the (Euclidean) finite $T$ correlators from their zero 
temperature counterparts by using the coordinate transformation from say the complex $w$-plane to the infinite cylinder with circumference $\beta$:
\be \label{transcy}
z\,=\,\frac{\beta}{2\pi}\,\ln\,w\,,\qquad z\,\equiv\sigma+i\tau\,,\qquad-\infty<\sigma<\infty\,,\quad 0\leq \tau<\beta\,.
\ee
Applying the transformation law for a (3,0) tensor, $W(w) \to 
\left(\tfrac{\partial w}{\partial z}\right)^3W(w(z))$, we obtain 
\be
\langle W(z_1)\,W(z_2) \rangle_{\mathbb R \times S^1_\beta}\,=\,{\cal N}\,\frac{\pi^6}{\beta^6\sinh^6\left(\tfrac{\pi}{\beta}(z_1-z_2)\right)}\,.
\ee
The normalization factor ${\cal N}$ depends on the central charge $c$, and for the currents in the Euclidean theory, it is negative. This is indeed the case for the operator product expansions (OPEs) and Ward identities deduced from the bulk Chern-Simons equations of motion \cite{Gutperle:2011kf, Ammon:2011nk, Kraus:2011ds}, wherein the leading singularity in the $WW$ OPE is \eqref{opew3}
\be
W(z)W(0)\,\sim\,-\frac{5\,c}{6\pi^2\,z^6}\,+\ldots
\label{opew}
\ee
 Adopting these conventions we have
\be \label{norm}
{\cal N}\,=\, (2\pi i)^{-2}\,\frac{10\,c}{3}\,.
\ee
With this identification, the chemical potential $\mu$ should match the corresponding object in the bulk gravity solution of \cite{Gutperle:2011kf}.
The prefactor of $(2\pi i)^{-2}$, as pointed out in Appendix D of  \cite{Gaberdiel:2012yb}, is natural when defining the zero modes or conserved charges on a spatial circle of size 
$2\pi$\footnote{Furthermore, the factor of $i^2$ can also be explained by recalling the well known fact that a chemical potential for global U(1) charge in the Euclidean Lagrangian formulation is equivalent to a constant {\em imaginary} background gauge field coupled to the conserved charge density. By analogy, the spin-three chemical potential deformation in the Euclidean theory should have a similar factor of $i$ included in the definition of the current $W$.}. Although we are working in the high temperature limit in which the spatial circle is effectively decompactified, we will continue to adopt the above conventions for ease of comparison to existing results.

We expect the one-point function for $W(z)+\overline W(\bar z)$ to vanish at the conformal point (with zero chemical potential), a fact that is easily verified for free boson and free fermion theories. Hence the first
correction to the partition function should appear at order $\mu^2$:
\bea
&&\frac{\ln Z}{L}\,=\,\frac{1}{L}\left[\ln Z^{(0)}_{\rm CFT}\,+
\right.\\\nonumber
&&\left.+\,\tfrac{1}{2}{\mu^2}\int_0^\beta d\tau_2\int_{-\infty}^\infty d\sigma_2
\int_0^\beta d\tau_1\int_{-\infty}^\infty d\sigma_1\,
\frac{{\cal N}\pi^6}{\beta^6\sinh^6\left(\tfrac{\pi}{\beta}(z_1-z_2)\right)}\,+\,{\rm h.c.}\ldots\right]\,,\\\nonumber
&&z_{1,2}\,\equiv\,\sigma_{1,2}+i\tau_{1,2}\,.
\eea 
Here $L$ is the spatial size of the system which is being taken to infinity; more precisely $\beta/L \to 0$. The integral can be analytically obtained using eq.\eqref{intthermal}. It is important to note that in order to arrive at the correct non-vanishing result, we perform integration along the non-compact spatial directions first.

A less direct, but useful method which sheds further light on the result is to note that the integrand, $\csch^6\left(\tfrac{\pi z}{\beta}\right)$, has a double pole at $z=0$ (in addition to higher order poles):
\be
\tilde z\,:=\,\frac{\pi z}{\beta}\,,\qquad\qquad\csch^6\tilde z \,=
\,\frac{1}{{\tilde z}^6}-\frac{1}{{\tilde z}^4}+\frac{8}{15 {\tilde z}^2} +\ldots\,.
\ee
When integrated, this yields a simple pole $\sim\,-\frac{8}{15\,\tilde z}$. `Trigonometrizing' the result, i.e. summing over the images of the simple pole shifted by integer multiples of $i\pi $ yields  $-\frac{8}{15}\coth\tilde z$
\footnote{This can be seen from the formal identity $\sum_{n=1}^\infty\left(\frac{1}{\tilde z-n\pi i} + \frac{1}{\tilde z+n\pi i}\right) + \frac{1}{\tilde z}\,=\,\coth\tilde z$.}. Evaluating this at $\tilde z \to \pm \infty$ and taking the difference between the two limits yields the desired answer 
$\int_{-\infty}^\infty d\tilde z\,\csch^6\tilde z\,=\,-\tfrac{16}{15}$. Higher order poles (along with their periodic images) produce exponentially decaying functions on the cylinder and do not contribute. Simple poles in the integrand, which are absent here, can also give rise to finite contributions  which will be important later.

It is also worth noting that the coefficient of the second order pole in the $W(z)W(0)$ OPE \eqref{opew3}, which determines the free energy correction, is proportional to the one-point function of the spin-4 current in the ${\cal W}$-algebra (or that of $:TT:$ for ${\cal W}_3$). 

 Thus, integration over $(\sigma_1,\tau_1)$  yields a constant independent of $(\sigma_2,\tau_2)$, and hence the second integration simply gives rise to an extensive scaling with the volume of the cylinder:
\be
\frac{1}{L}\ln Z\,=\,
%\frac{1}{L}\ln Z^{(0)}_{\rm CFT}\,+\,\tfrac{8\pi^3}{9}\,c\,\mu^2\,\beta^{-3}\,+\ldots
\frac{\pi\,c}{6}\,\beta^{-1}\,+\,\frac{8\pi^3\,c}{9}\,\mu^2\,\beta^{-3}\,+\ldots\,.
\label{hightz}
\ee
In terms of the variables $\tau\,\equiv\,i\beta/(2\pi)$ and $\alpha\,\equiv\,-i\mu\beta/(2\pi)$, the expression can be rewritten as
\be
\frac{2\pi}{L}\ln Z\,=\,\frac{i\pi^2 c}{6\,\tau}\,\left(1-\frac{4}{3}\frac{\alpha^2}{\tau^4}\,+\,\ldots\right)\,.\label{thermalz}
\ee
This precisely matches both the first correction to the high temperature partition function of the spin-three black hole of Gutperle and Kraus \cite{Gutperle:2011kf} and the CFT computations of 
\cite{Gaberdiel:2012yb, Kraus:2011ds}. We recall that \cite{Gaberdiel:2012yb} utilized the modular transformation properties of torus amplitudes, taking their high temperature limits and evaluating the resulting integrals. The work of \cite{Kraus:2011ds} took advantage of the free field realization of ${\cal W}_{\infty}[\lambda]$ symmetry for $\lambda=0,1$ and computed the exact thermal partition function.
Here we have employed conformal perturbation theory, directly used the two-point correlator  of $W$-currents, and calculated its integral on the cylinder ${\mathbb R}\times S^1_\beta$. It is clear that the strategy can be employed to deduce the ${\cal O}(\mu^4)$ correction using the 4-point function of $W$-currents on the cylinder. We leave this for future study. Below, we will adopt the same approach to compute the entanglement entropies at finite temperature for free field realizations of CFTs with ${\cal W}$-symmetry.

A notable feature of the free energy correction at ${\cal O}(\mu^2)$ is that it is universal, relying only on the existence of the spin-three current and not on the details of the ${\cal W}$-algebra associated to the CFT.

\section{Entanglement entropy and conformal perturbation theory }
\label{pertEE}
The entanglement entropy for a subsystem $A$ is defined in terms of its reduced density matrix $\rho_A$, as the associated von Neumann entropy:
\be
S_A\,=\,-{\rm Tr}\,\rho_A\ln\rho_A\,.
\ee
The direct evaluation of the reduced density matrix and the von Neumann entropy  invariably presents a challenging task.  Instead, motivated by the replica trick, one may obtain the entanglement entropy via an indirect route, by first calculating ${\rm Tr}\rho^n$ for any $n\geq 1$ and then examining the limit \cite{Calabrese:2004eu, Calabrese:2009qy},
\be
S_A\,=\, -\lim_{n\to1}\,\frac{\partial}{\partial n}\,{\rm Tr}\rho_A^n\,.
\ee
In this approach it is customary to define the entanglement entropy of subsystem $A$ as a limit of the  R\'enyi entropies $S_A^{\,(n)}$,
\be
S_A^{\,(n)}\,=\,\frac{1}{1-n}\,\ln\,{\rm Tr}\rho_A^n\,,\qquad\qquad
S_A \,=\, \lim_{n\to 1}S_A^{\,(n)}\,.
\ee
For the entanglement entropy of a single interval in a CFT$_2$, the $n$-replicated
partition function is given by the partition function on an $n$-sheeted Riemann surface with two branch points $y_1$ and $y_2$, where the latter represent the positions of the end-points of the interval:
\begin{align}
\tr  \rho ^n_A = \frac{Z^{\,\n}_A}{(Z)^n} 
\end{align}
Here $Z^{\,(n)}_A$ is the CFT partition function on the $n$-sheeted Riemann surface, whilst $Z$ is the standard CFT partition function on the single sheeted Riemann  surface. The partition function on the multi-sheeted Riemann surface is then computed by introducing $n$ replica copies/branches of fields $\{\varphi_i\}_{i=1,\ldots,n}$ on the complex plane (taking $\varphi$ to be a free boson, for instance), with the copies being cyclically permuted across successive Reimann sheets glued along the branch cut. The action of this cyclic permutation can be diagonalized by considering the linear combinations
\be
\tilde\varphi_k\,=\,\sum_{\ell=0}^{n-1}\,e^{2\pi i \ell\frac{k}{n}}\,\varphi_\ell\,,\qquad k\,=\,0,1,\ldots n-1\,.
\ee
Upon going around a branch point $y_1$ (or $y_2$) the field $\tilde \varphi_k$ gets multiplied by a phase factor $e^{2\pi i k/n}$ ($e^{- 2\pi i k/n}$), which can be viewed as being due to a twist (anti-twist) field insertion at the location of the branch point in the $k$th sheet. The partition function on the $n$-sheeted Riemann surface can therefore be expressed as a correlation function of $n$-pairs of twist-anti-twist operators 
\begin{align}
Z^{(n)}\, =\, \prod_{k=0}^{n-1} \langle\, \sigma_{k,n} (y_1,\bar y_1)\,\, \overline{\sigma} _{k,n} (y_2,\bar y_2) \,  \rangle\,,
\end{align}
where the index $k$ corresponds to the $k$th Riemann sheet. By translational invariance, the result for the entanglement entropy should only be a function of the interval length,
\be
\Delta\,\equiv\, |y_1-y_2|\,.
\ee

Our aim is to calculate this twist correlator in a CFT with ${\cal W}$-algebra symmetry, deformed by a spin-three chemical potential. Specifically, we will be interested in the ${\cal O}(\mu^2)$ correction to the entanglement entropy, which will be determined using conformal perturbation theory at high temperature.
 We will focus attention on specific examples where the ${\cal W}$-symmetry is realized in terms of free fields (both bosons and fermions). However, we will uncover a result that is likely universal, independent of the details of the theory, much like the order $\mu^2$ correction to the thermal entropy/free energy deduced above.

The dependence of  entanglement/R\'{e}nyi entropy  of a CFT deformed by  operators have been 
studied earlier in \cite{CardyCalabrese,Sierra}.  These works dealt with perturbing operators which were
primaries with weights $(h, \bar h)$ where 
$\bar h \neq 0$. Thus unlike the case considered in this paper the 
 conformal perturbation theory is not holomorphic. 
 Holomorphic perturbation theory as we will show in this paper is considerably easier 
 to handle and exact results can be obtained. 
 Another important difference is that the perturbing operator considered in this paper 
 is a perturbation of the basic Lagrangian of the theory. Therefore when evaluating the 
 entanglement entropy using the replica trick, the perturbing operator is present in each 
 copy of the CFT which implies these operators are singlets under the cyclic permutation of
 the copies. This is unlike the situation considered in \cite{CardyCalabrese} where the operators were not 
 singlets under cyclic permutation of the replica copies.
 In \cite{Sierra}  the perturbation involved an insertion of $2n$ primaries whereas 
 we have an insertion of only $2$ primaries. 

\subsection{Entanglement entropy at ${\cal O}(\mu^2)$}
Expanding out the $n$-replicated partition function for a CFT perturbed by a chemical potential for spin-three charge, the 
R\'enyi entropy at order $\mu^2$ is given by a four point correlator involving two $W$-currents and a pair of twist-anti-twist fields:
\bea
&&S^{\,(n)}(\Delta)\,=\,\frac{1}{1-n}\,\ln \frac{1}{Z^n}\prod_a\prod_{k}\left( \langle\, \sigma_{k,n}^a (y_1,\bar y_1)\, \overline{\sigma}_{k,n}^a (y_2, \bar y_2) \,  \rangle_{\rm CFT}\,+\frac{\mu^2}{2}
\int d^2z_1\int d^2 z_2\right.\label{renyi}\nonumber\\
&&\left.\left\langle\, \sigma_{k,n}^a (y_1,\bar y_1)\, 
\left[W(z_1)+\overline W(\bar z_1)\right]\,\left[W(z_2)+\overline W(\bar z_2)\right]\,\overline{\sigma}_{k,n}^a (y_2,\bar y_2) \,  \right\rangle_{\rm CFT} +\ldots\right)
\eea
Here, $a$ is a species index and counts the number of free bosons or fermions in the theory. We have omitted the term linear in $\mu$ without {\em a priori} justification, but in the explicit examples we will study, the correlator involving a single insertion of $W+\overline W$ will vanish identically.

The partition function $Z$ of the deformed CFT has a high temperature expansion given by eq.\eqref{hightz}, so that
\be
(Z)^n\,\simeq\,e^{nL{\pi^2 c}/{6\beta}}\left(1\,+\,n\frac{8\pi^3\,c}{9}\,\frac{\mu^2}{\beta^2}\,\frac{L}{\beta}\,+\ldots\right)\,,\label{ztothen}
\ee
where $L$ is the size of the system, with $(L/\beta)\gg1$. We will see that a similar extensive term appears in the replicated partition function $Z^{(n)}_A$, which is precisely cancelled by the order $\mu^2$ correction to $Z^n$ in \eqref{ztothen}. 

Making use of translational invariance (along the non-compact spatial direction), without loss of generality we take
\be
y_1\,=\,0\,,\qquad  y_2 \,=\, \Delta\, \in {\mathbb R}\,.
\ee
The leading term in the R\'enyi entropy \eqref{renyi} arises from the two-point function of twist-operators, and in the limit $n\to 1$ yields the well known universal formula for EE of CFTs in two dimensions \cite{Calabrese:2004eu, Calabrese:2009qy, Casini:2009sr}, namely $S_{\rm EE}\,=\,c/3 \,\ln\Delta$ at zero temperature.

Whilst the correlators of the higher spin currents  $\langle WW\rangle$,  and the twist fields $\langle\sigma \bar\sigma\rangle$, are separately clearly known, the OPEs and correlation functions involving the twist fields along with  higher spin currents ($\langle\sigma\bar\sigma W\rangle$, $\langle\sigma \bar\sigma WW\rangle$) are not {\it a priori} obvious. For this reason we need to focus attention on specific examples wherein the four-point function entering the ${\cal O}(\mu^2)$ correction to the entanglement entropy can be determined by a direct calculation. The tractable examples are presented by free field realizations of ${\cal W}$-algebra symmetries and the holographic proposal of \cite{deBoer:2013vca}:
\begin{itemize}
\item{The theory of $N$ free fermions with ${\cal W}_{1+\infty}$ symmetry 
\cite{Bergshoeff:1990yd}, $\lambda=0$ and central charge $c=N$. This theory also has a spin-1 current which is absent in theories of higher spin gravity that are dual to ${\cal W}$-algebra CFTs in 2D.}
\item{The theory of $N$ complex free bosons with central charge $c=2N$, which has the infinite dimensional  ${\cal W}_{\infty}[\lambda]$ symmetry with $\lambda=1$ \cite{Pope:1989ew, Kraus:2011ds}. }
%\item{The theory obtained by cosetting with respect to the U(1) current in the theory of $N$ free fermions \cite{Gaberdiel:2013jpa}. We find that the procedure of cosetting furnishes a free field representation of finite ${\cal W}_N$ algebras. }
\item{The proposed holographic entanglement entropy for higher spin theories of gravity on AdS$_3$ \cite{deBoer:2013vca, Ammon:2013hba} allows to compute EE for higher spin black holes within the Chern-Simons formulation. The work of \cite{deBoer:2013vca} also includes a proposal for the so-called holomorphic EE, for which explicit results only exist in the case with ${\cal W}_3$ symmetry.}
\end{itemize}

\subsection{Universal result for ${\cal O}(\mu^2)$ EE at finite temperature}

For all three classes of theories above, we find by explicit computation a universal, non-trivial formula for the order $\mu^2$ correction to entanglement entropy as a function of temperature $T=\beta^{-1}$ and interval length $\Delta$. We defer the details of the calculations for each specific example to subsequent sections. At this stage, we find it useful to summarize and present the outcome of these calculations and the final result.

By evaluating the correlators involving twists and higher spin currents and accounting for appropriate combinatorial factors, we find, upon taking the $n\to 1$ limit for the R\'enyi entropy $S^{\,(n)}$,
\bea
\boxed{S_{\rm EE}(\Delta)\,=\,\frac{c}{3}\,\log\Big|\frac{\pi}{\beta}\sinh\left(\frac{\pi\Delta}{\beta}\right)\Big|\,-\,\mu^2{\cal N}\,\left[\,\frac{1}{2}\,{\cal I}_1\left(\Delta\right)\,-\,\frac{1}{20}\,{\cal I}_2\left(\Delta\right) \right]\,+\,{\cal O}(\mu^4).}\nonumber\\
\label{EEmain}
\eea
Here ${\cal N}= - \frac{5c}{6\pi^2}$ as in \eqref{norm} and  ${\cal I}_1$ and ${\cal I}_2$ are two types of integrals contributing to the entanglement entropy, originating from the operator products within the  four-point function $\langle\sigma\bar\sigma WW\rangle$ on the cylinder ${\mathbb R}\times S^1_\beta$,
\bea
&&{\cal I}_1\left(\Delta\right)\,=\,\int d^2z_1\int d^2 z_2
\,\,H^4(z_1-z_2)\,G (z_1)\,G(z_2)\label{I1}\\\nonumber\\\nonumber
&& {\cal I}_2\left(\Delta\right)\,=\,\int d^2z_1\int d^2 z_2
\,\,H^2(z_1-z_2)\,G^2 (z_1)\,G^2(z_2)\label{I2}
\eea
where 
\be
H(z)\,\equiv\,\frac{\pi}{\beta \sinh\left(\frac{\pi\,z}{\beta}\right)}\,,\qquad\qquad
G(z)\,\equiv\,\frac{\pi\,{\sinh\left(\frac{\pi\,\Delta}{\beta}\right)}}{\beta \sinh\left(\frac{\pi\,z}{\beta}\right)\,\sinh\left(\frac{\pi\,(z-\Delta)}{\beta}\right)}\,.\label{HGdef}
\ee
Technically, the origin of these two types of terms is different -- the integrand in ${\cal I}_1$ arises from simple poles in the OPEs of twist operators with the $W$-currents, whilst ${\cal I}_2$ originates from double pole singularities in the same. 
Fortunately, the integrals on the cylinder, while somewhat tedious to evaluate (see appendix \ref{integrals} for details), are analytically tractable and yield compact final expressions:
\bea
&&{\cal I}_1\left(\Delta\right)\,=\, \frac{4\pi^4}{3\beta^2}\,\left(\frac{4\pi\Delta}{\beta}\,\coth\left(\tfrac{\pi\Delta}{\beta}\right)\,-\,1\right)\,+\,
\label{I1final}\\\nonumber
&&\qquad\qquad+\,\frac{4\pi^4}{\beta^2}\sinh^{-2}\left(\tfrac{\pi\Delta}{\beta}\right)\,\left\{\left(1-\frac{\pi\Delta}{\beta}\coth\left(\tfrac{\pi\Delta}{\beta}\right)\right)^2\,-\,\left(\tfrac{\pi\Delta}{\beta}\right)^2\right\}
\eea
and
\bea
&&{\cal I}_2\left(\Delta\right)\,=\,\frac{8\pi^4}{\beta^2}\,\left(5\,-\,\frac{4\pi\Delta}{\beta}\,\coth\left(\tfrac{\pi\Delta}{\beta}\right)\right)\,+\,
\label{I2final}\\\nonumber
&&\qquad\qquad+\,\frac{72\pi^4}{\beta^2}\sinh^{-2}\left(\tfrac{\pi\Delta}{\beta}\right)\,\left\{\left(1-\frac{\pi\Delta}{\beta}\coth\left(\tfrac{\pi\Delta}{\beta}\right)\right)^2\,-\,\frac{1}{9}\left(\tfrac{\pi\Delta}{\beta}\right)^2\right\}\,.
\eea
The only subtle point here is that one must perform the integration over the non-compact coordinates first, following which the integrals over the periodic coordinates are rendered trivial. In appendix \ref{integrals} we have explained in detail how these integrals are performed by examining singularities of the integrand, following the basic idea outlined in 
section \ref{holpert}.

We list below the important features of and checks satisfied by the expression
\eqref{EEmain}.
\begin{itemize}
\item{The first noteworthy feature which will be established below by explicit computation, is that \eqref{EEmain} holds for both the free fermion and free boson CFTs deformed by the spin-three chemical potential.}
\item{In the high temperature limit, $(T\Delta)\to \infty$, it becomes  purely extensive with respect to the interval length $\Delta$ and reproduces the thermal entropy that follows from eq.\eqref{thermalz}
\be
S_{\rm EE}\big|_{(\beta/\Delta)\ll1}\to\frac{c}{3}\pi T\,\Delta\,+\,
\frac{32\pi^3}{9}\,(\mu T)^2\,T\Delta\,+\ldots
\ee
Note that the validity of the perturbative expansion simultaneously requires $\mu^2 T^2 \ll 1$. The subleading terms include exponentially suppressed corrections at high temperature.
}
\item{At low temperatures $(T\Delta)\to 0$, the correction term at order $\mu^2$ is vanishing,
\be
S_{\rm EE}\big|_{(\beta/\Delta)\gg 1}\to \frac{c}{3}\ln\Delta\,+\,\frac{40\pi^4}{27}\,(\mu T)^2\,T^2\Delta^2\,+\ldots
\ee
This is in stark contrast to the behaviour of the proposed entanglement entropy in the {\em canonical} formalism \cite{deBoer:2013vca}, for the planar spin-three black hole in ${\rm SL}(3,\mathbb R)\times{\rm SL}(3, \mathbb R)$ Chern-Simons theory. 
 }
\item{Most remarkably, the formula \eqref{EEmain} precisely matches the proposal of de Boer and Jottar \cite{deBoer:2013vca} for the {\em holomorphic} entanglement entropy of the spin-three planar black hole in the bulk ${\rm SL}(3,{\mathbb R})\times{\rm SL}(3,{\mathbb R})$ Chern-Simons theory. We will review their proposal and expand upon this surprising agreement below.}

\item{Finally, the agreement with the bulk gravity proposal and the fact that both free boson and free fermion CFTs with ${\cal W}$-algebra symmetries yield the same result, supports the universal ($\lambda$-independent) nature of this particular correction.}
\end{itemize}

\section{Free fermions: ${\cal W}_{1+\infty}$ algebra at $\lambda=0$}
\label{freefer}
In this section we will present the nuts and bolts of the calculation leading to the result \eqref{EEmain} in the theory with $N$ free fermions. This is the simplest of the three examples we will look at. The free fermion theory provides a realization of the ${\cal W}_{1+\infty}$ algebra \cite{Bergshoeff:1990yd, Pope:1991ig} which contains currents of all spins $s\geq 1$.

 The theory of $N$ free complex fermions has central charge $c=N$, and possesses a $U(N)$ global symmetry under which the fermions $\{\psi^a\}_{a=1,\ldots N}$ transform in the fundamental representation. The OPE for the free fermions is
\be
\psi_a^*(w_1)\,\psi^b(w_2)\,\sim\,\frac{\delta_a^b}{w_1-w_2}\,.
\ee
Since $\psi^a$ and $\psi_a^*$ are $(\frac{1}{2},0)$ tensors we can use their transformation property to write down their correlator on the finite temperature Euclidean cylinder ${\mathbb R}\times S^1_\beta$, using $z(w)\,=\,\frac{\beta}{2\pi}\ln w$:
\be
\langle\psi^*_a(z_1)\psi^b(z_2)\rangle_{{\mathbb R}\times S^1_\beta}\,=\,\frac{\pi\,\delta_a^b}{\beta\sinh\left(\frac{\pi}{\beta}\left(z_1-z_2\right)\right)}\,.
\ee
The theory has an infinite set of higher spin conserved currents in the $U(N)$-singlet sector. The first few of these, the spin-1, spin-2 (the stress tensor) and the spin-3 currents are
\bea
&&J\,=\,\psi_a^*\psi^a\,,\qquad\qquad T\,=\,\tfrac{1}{2}\left(\partial\bar\psi_a^*\psi^a-\psi_a^*\partial\psi^a\right)\,,\\
&& W\,=\,i\frac{\sqrt 5}{12\pi}\left(\partial^2\psi_a^*\psi^a \,- \,4\,\partial\psi_a^*\partial\psi^a + \psi_a^*\partial^2\psi^a\right)\,.\label{fermionspin3}
\eea
We have normalized the spin-three current as in \cite{Kraus:2011ds} to reproduce the leading singularity in the OPE \eqref{opew}. As it stands, the spin-three current $W$ is not a primary field in this theory because its OPE with the stress tensor contains a term proportional to the spin-1 current $J$.

As we have remarked earlier, the holographic duality proposal due to Gopakumar and Gaberdiel  relates ${\cal W}$-algebra CFTs to higher spin theories on AdS$_3$ with an asymptotic ${\cal W}_\infty[\lambda]$ symmetry algebra. There are two possible ways of restricting ${\cal W}_{1+\infty}[0]$ to ${\cal W}_\infty[0]$, which have been discussed in the literature. The first method is to introduce a chemical potential for $J$ to fix the U(1) charge to zero as was done in the partition function computation of \cite{Kraus:2011ds}. The second approach is to perform a cosetting of the algebra by the U(1)-current $J$, as discussed by \cite{Gaberdiel:2013jpa}. We will return to both these points subsequently, but for now we will continue to work with the naive definition of the spin-three current \eqref{fermionspin3} and introduce a chemical potential for it.

\subsection{Bosonization}
Computing the R\'enyi entropy for fermions requires a slight modification of the action of twist operators on the elementary fields discussed in section \ref{pertEE}, due to Fermi statistics. The fermion replica field on the $n$-th sheet, upon crossing the branch cut, gets identified with replica field on the first sheet up to a minus sign, depending on whether $n$ is even or odd \cite{Calabrese:2004eu, Casini:2009sr}. Diagonalizing the action of twist operators yields a multiplicative  phase factor ${e^{2\pi i k/n}}$, $k= -\frac{1}{2}(n-1)\,,-\frac{1}{2}(n-3)\,\ldots ,\frac{1}{2}(n-1)$, on the (diagonalized) replica fields, upon going around a branch point. This fixes the action of the twist operators on the free fermions.

To find an explicit representation of the twist fields for free fermions, we move to the bosonized language so that
\be
\psi^{k,a}(z)\,=\,:e^{i\varphi_{a,k}(z)}:\,\qquad
\psi^{*k}_a(z)\,=\,:e^{-i\varphi_{a,k}(z)}:\,
\ee
and similarly for the anti-holomorphic sector. Here $\{\varphi_{a}(z)\}$ are free (chiral) bosons with the OPE,
\be
\varphi_{a}(z)\varphi_{b}(0)\,\sim\, -\,\delta_{a,b}\,\ln(z)\,.
\ee
The twist fields for each $k\,=\, -\tfrac{(n-1)}{2},-\tfrac{(n-3)}{2}\ldots \tfrac{(n-1)}{2}$, can then be represented as,
\be
\sigma_{k,n}(z,\bar z)\,=\,\prod_{a=1}^N :e^{i\frac{k}{n}\left(\varphi_{a,k}(z)-\bar\varphi_{a,k}(\bar z)\right)}:\qquad
\bar\sigma_{k,n}(z,\bar z)\,=\,\prod_{a=1}^N :e^{-i\frac{k}{n}\left(\varphi_{a,k}(z)-\bar\varphi_{a,k}(\bar z)\right)}:\nonumber
\ee
The conformal dimension of the vertex operator $:e^{i\frac{k}{n}\varphi_{a,k}}:$ is $k^2/(2n^2)$.
Hence the product of  twist fields over all allowed values of $k$ yields the branch point twist field associated to the $n$-sheeted Riemann surface, whose  dimension $\Delta_n\,=\,\frac{N}{24}\left(n-\tfrac{1}{n}\right)$. 
Using the standard OPE between vertex operators, we can see that the twist fields generate precisely the branch cuts necessary to produce the correct phases upon taking the replica fermions around the branch point.

To avoid cluttering formulae below, we will suppress the replica label ($k$) on the fields unless it is necessary for clarity.

In the bosonized language, the U(1) current $J$, the stress tensor $T$ and spin-three current $W$ become (see e.g. \cite{Pope:1991ig})
\bea
J\,=\,i\sum_{a=1}^N\partial\varphi_a\,,\qquad T\,=\,-\tfrac{1}{2}\sum_{a=1}^N:(\partial\varphi_a)^2:\,,\qquad
W\,=\,-\frac{\sqrt{5}}{6\pi}\sum_{a=1}^N
:(\partial\varphi_a)^3:\label{wbosonized}
\eea
Once again, we have fixed the normalization of $W$ so that it matches the leading singularity of the OPE \eqref{opew}.

\subsection{R\'enyi and entanglement entropies at ${\cal O}(\mu^2)$}
\label{EEREff}
Let us now consider the R\'enyi entropy of the free fermion CFT with  spin-three chemical potential as in eq.\eqref{action}. We will expand the action up to quadratic order in $\mu$, treating the chemical potential as a perturbation. Following the discussion in section \ref{pertEE} it is now clear that for this calculation we need the one- and two-point functions of $(W+\overline W)$ in the presence of the twist operators. 

The calculation of these can be simply performed by Wick contractions of the free boson fields $\varphi_{a}$, using the two-point function on the 
cylinder ${\mathbb R}\times S^1_\beta$
\be
\langle\varphi_{a}(z_1)\,\varphi_{b}(z_2)\rangle\,=\,-\,\delta_{a,b}\,\ln\,\sinh\left(\tfrac{\pi}{\beta}(z_1-z_2)\right)\,.\label{2ptcylinder}
\ee
For a holomorphic tensor operator $O_h$with conformal dimension $h$, the two-point function  on the plane  can be transformed to the cylinder ${\mathbb R}\times S^1_\beta$ to yield, 
\be
\langle O_h(z_1)\,O_h(z_2)\rangle\,=\,
\frac{\pi^{2h}}{\beta^{2h}\sinh\left(\tfrac{\pi}{\beta}(z_1-z_2)\right)^{2h}}\,.
\ee 
Our strategy will be to first rewrite the operator product of the twist and anti-twist fields as a normal ordered product,
\be
\sigma_{k,n}(y_1,\bar y_1)\,\bar\sigma_{k,n}(y_2,\bar y_2)\,=\, \frac{:\sigma_{k,n}(y_1,\bar y_1)\,\,\bar\sigma_{k,n}(y_2,\bar y_2):}{\big|\frac{\beta}{\pi}\sinh\left(\frac{\pi\Delta}{\beta}\right)\big|^{2k^2\,N/n^2}}.\label{tta}
\ee
In addition, insertions of a pair of $W$-currents (with $W\sim\, :(\partial\varphi_a)^3:$)  can be expressed in terms of normal ordered products,
\bea
&&\sum_{a,b=1}^N:(\partial\varphi_a(z_1))^3:\,:(\partial\varphi_b(z_2))^3:\,=\label{wwope}
\\\nonumber
&&-\,6N\,H^6(z_1-z_2)\,-\,9\,\sum_{a=1}^N:(\partial\varphi_a(z_1))^2(\partial\varphi_a(z_2))^2: 
H^2(z_1-z_2)\nonumber\\\nonumber
&&+18 \sum_{a=1}^N:(\partial\varphi_a(z_1))(\partial\varphi_a(z_2)): H^4(z_1-z_2)\,+
\sum_{a,b=1}^N:(\partial\varphi_a(z_1))^3(\partial\varphi_b(z_2))^3:
\eea
where $H(z_1-z_2)$ is defined in eq.\eqref{HGdef} and determines the two-point function $\langle\partial\varphi_a(z)\partial\varphi_b(0)\rangle=\,-\,
\delta_{a,b}\,H^2(z)$, on the  cylinder. This is an important step. We are effectively taking the two-point function of $W$ on the cylinder to be 
\be
\langle W(z_1)\,W(z_2)\rangle\,=\,-\frac{5N}{6\pi^2}\frac{\pi^6}{\beta^6\sinh^6\left(\tfrac{\beta}{\pi}(z_1-z_2)\right)}.
\label{2pointw}
\ee
Thus we have implicitly set $J=0$, so that the $W$-current transforms as a $(3,0)$ tensor.

Finally, we will make repeated use of  the operator product between $\partial\varphi_{a,k}$ and the normal ordered twist operators on the cylinder,
\bea
&&\partial\varphi_{a,k}(z)\,:\sigma_{k,n}(y_1, \bar y_1)\,\bar\sigma_{k,n}(y_2, \bar y_2):
\,\,=\,\left(\frac{ik}{n}\right)\,G(z)\times
\label{opeG}\\\nonumber
&&\hspace{2in}:\sigma_{k,n}(y_1, \bar y_1)\,\bar\sigma_{k,n}(y_2, \bar y_2):\,+\,\,{\rm normal\, ordered}
\eea
where $G(z)$ was defined previously in eq.\eqref{HGdef} with $y_1=0$ and $y_2=\Delta \in {\mathbb R}$.
This is deduced by applying the standard rules for operator products of free fields and the two-point function eq.\eqref{2ptcylinder}. The ``normal ordered'' terms above refer to regular pieces which annihilate the vacuum. The anti-holomorphic operator $\bar\partial\bar\varphi_{a,k}(\bar z)$ has a very similar operator product with the twist-anti-twist pair, but with the opposite sign $\bar\partial\bar\varphi_{a,k}(\bar z):\sigma_{k,n}\,\bar\sigma_{k,n}:\, =\, -(ik/n)G(\bar z):\sigma_{k,n}\,\bar\sigma_{k,n}: + \ldots$.

\paragraph{{First order correction in $\mu$}:} Taking $y_1=0$ and $y_2 =\Delta \in {\mathbb R}$,
we begin with the putative correction at order $\mu$, using the bosonized form \eqref{wbosonized} for the spin-three current,
\bea
&&-\mu\int d^2 z\,\langle\sigma_{k,n}(y_1,\bar y_1) \,\bar\sigma_{k,n}(y_2,\bar y_2) 
\left(W(z)\,+\,\overline W(\bar z)\right)\rangle
\label{ff1pt}\\\nonumber
&&\qquad\qquad=\,\frac{\sqrt 5\,\mu N}{6\pi}\,\left(\frac{ik}{n}
\right)^3{{\Big|\frac{\beta}{\pi}\sinh\left(\frac{\pi\Delta}{\beta}\right)\Big|^{-2k^2\,N/n^2}}} \int d^2 z
\left[G(z)^3\,-\,G(\bar z)^3
\right]\,=\,0\,.
\eea
The expression was obtained by using eq.\eqref{tta}
and three-fold application of \eqref{opeG}, in order to evaluate $\langle:(\partial\varphi_{a,k})^3:\sigma_{k,n}\bar\sigma_{k,n}\rangle$.
The end-points of the interval $(y_1,y_2)$ have been taken to be real  so that they are associated to a {\em spatial} interval on the cylinder ${\mathbb R}\times S^1_\beta$. With this choice the two integrals cancel against each other, as can be seen immediately by performing the variable change 
$\tilde z \,=\bar z$ in the second 
term. The cancellation occurs because we have chosen equal chemical potentials for both holomorphic and anti-holomorphic sectors. 

To show that the first order correction in $\mu$ vanished we used the 
fact that chemical potentials were equal and the integrals involved cancel. 
In fact the vanishing of the linear term in $\mu$ is a more stronger result and it occurs
when $\mu \neq- \bar \mu$. 
This is because the one point function 
\bea \label{onept}
&&\int d^2 z\,\langle
\prod_{-\frac{n-1}{2} }^{\frac{n-1}{2}} \sigma_{k,n}(y_1,\bar y_1) \,\bar\sigma_{k,n}(y_2,\bar y_2) W(z)\rangle\,=\, 
\sum_{-\frac{n-1}{2}}^{\frac{n-1}{2}} \frac{\sqrt 5\, N}{6\pi}\,\left(\frac{ik}{n}
\right)^3
\frac{4\pi^2}{\beta} \\ \nonumber
& & \hspace{1.2in}\times \left[3\coth\left(\tfrac{\pi\Delta}{\beta}\right)\,+\,\frac{\pi\Delta}{\beta}\left(1-3\coth^2\left(\tfrac{\pi\Delta}{\beta}\right)\right)\right]
{{\Big|\frac{\beta}{\pi}\sinh\left(\frac{\pi\Delta}{\beta}\right)\Big|^{-\Delta_n }}}
\,,
\eea
vanishes on summing $k$ from $-\frac{1}{2} (n-1) $ to $\frac{1}{2}( n-1) $. 
In the above equation we have used the fact that  $W$ is the sum of the currents from each of the 
Fourier copies. 
A more general way to see this result is that the one point function 
$\langle W(z) \prod_k \sigma_k \bar \sigma_k \rangle$  is equal to  
the one point function $(w'(z))^3\langle W(w(z)) \rangle_{{\cal C}}$ where 
${\cal C}$ is the complex plane
obtained by uniformization of the  $n$-sheeted  Riemann surface branched along the interval
between the two branch point twist operators.  
$w(z)$ is the conformal transformation which takes the $n$-sheeted Riemann surface
to the complex plane. We have also used that 
 in the charge $J=0$ sector the spin-3 current $W$ transforms as a primary. 
  Now by Lorentz invariance
$\langle W(w(z)) \rangle_{{\cal C}}$ vanishes and therefore the three point function 
$\langle W(z) \prod_k \sigma_k \bar \sigma_k \rangle$  vanishes. 
The same argument ensures that the three point function involving
 the anti-holomorphic current $\bar W$ and the branch point twist operators vanish. 
This argument is analogous to the argument used by \cite{Calabrese:2004eu} to evaluate the 
three point function $\langle T(z) \prod_k \sigma_k \bar \sigma_k \rangle$  where 
$T$ is the stress energy tensor of the CFT. 
Thus we conclude that the linear term in $\mu, \bar \mu$ vanishes even if
$\mu\neq- \bar \mu$.

\paragraph{{Second order correction}:}
We now have all the ingredients in place to write down the order $\mu^2$ contribution, including the situation with $\mu\neq - \bar \mu$.  However, the static case  
$\mu= -\bar\mu$ is our primary focus for now and we will outline the calculation for this explicitly.

The four point correlator that determines the correction at this order can be written down quite straightforwardly using eqs.\eqref{tta}, \eqref{wwope} and \eqref{opeG}:
\bea
&&\tfrac{1}{2}\int d^2z_1\int d^2z_2
\left\langle\, \sigma_{k,n} (y_1,\bar y_1)\, 
\left[W(z_1)+\overline W(\bar z_1)\right]\,\left[W(z_2)+\overline W(\bar z_2)\right]\,\overline{\sigma}_{k,n} (y_2,\bar y_2) \,  \right\rangle_{\rm CFT}
\nonumber\\\nonumber\\
&&=\,\frac{5N}{6\pi^2}\,\Big|\frac{\beta}{\pi}\sinh\left(\frac{\pi\Delta}{\beta}\right)\Big|^{-2k^2\,N/n^2}\int d^2z_1\int d^2z_2\ \Bigg(-\,H^6(z_1-z_2)\,
 \label{4ptff}\\\nonumber\\\nonumber
&& \qquad \quad +\,  3 \left(\frac{ik}{n}\right)^2\,
H^4(z_1-z_2)G(z_1)G(z_2)\,-\,\frac{3}{2}\left(\frac{ik}{n}\right)^4\,H^2(z_1-z_2)G^2(z_1)G^2(z_2)\Bigg)
\eea
We have taken into account both holomorphic and anti-holomorphic contributions. In doing so, we also observe an exact cancellation between terms arising from within
$\langle:\sigma\bar\sigma:\,:(W+\overline W)(W+\overline W):\rangle$.

The first term in eq.\eqref{4ptff}, proportional to $\int\int H^6(z_1-z_2)$, is precisely the order $\mu^2$ correction to the thermal partition function which is cancelled by the normalization $Z^n$ in the  R\'enyi entropy 
\eqref{renyi}. The R\'enyi entropy may be written as a sum over replica sectors labelled by $k$ \footnote{
The partition function $Z_k$ at order $\mu^2$ is
\bea
&&
Z_k\,=\,\Big|\tfrac{\beta}{\pi}\sinh\left(\tfrac{\pi\Delta}{\beta}\right)\Big|^{-2k^2 N/n^2}\left[1\,+
\,N\,\tfrac{8\pi^3}{9}\tfrac{\mu^2L}{\beta^3}\,+\,
\tfrac{5N\mu^2}{6\pi^2}
\left(
3 \left(\tfrac{ik}{n}\right)^2\,{\cal I}_1(\Delta)\,-\,\tfrac{3}{2}\left(\tfrac{ik}{n}\right)^4\,{\cal I}_2(\Delta)
\right)+\ldots\right]\,.\nonumber
\eea
}
\be
S^{\,(n)}\,=\,\frac{1}{1-n}\sum_{k}\,\ln\left(\frac{Z_{k}}{Z}\right)\,,
\qquad\quad k\,=\,- \tfrac{n-1}{2}, \ldots,\tfrac{n-3}{2}, \tfrac{n-1}{2}\,.
\ee
Making use of the sums over $k$,
\be
\sum_{k}\left(\frac{ik}{n}\right)^2
\,=\,\frac{1-n^2}{12\,n}\,,\qquad\qquad
\sum_{k}\left(\frac{ik}{n}\right)^4
\,=\,\frac{(1-n^2)(7-3n^2)}{240\,n^3}\,,
\ee
and utilizing the shorthand notation for the integrals as in 
eqs.\eqref{I1} and \eqref{I2}, 
we find that the R\'enyi entropy at this order is
\bea\label{renyiff}
&&S^{\,(n)}(\Delta)\,=\,\frac{N}{6}(n+1)\ln\Big|\frac{\beta}{\pi}\sinh\left(\frac{\pi\Delta}{\beta}\right)\Big|\,\\\nonumber\\\nonumber
&&\hspace{1.5in}+\,\frac{5N\mu^2}{6\pi^2}\left(\frac{(1+n)}{4n}\,{\cal I}_1(\Delta)\,-\,\frac{(1+n)(7-3n^2)}{160n^3}\,{\cal I}_2(\Delta)\right)\,+\ldots
\eea
Eventually, we will find it interesting to compare the above result for R\'enyi entropies in the free fermion CFT with the corresponding formula for free bosons. Finally, we take the limit $n\to 1$ of the 
R\'enyi entropy to arrive at the entanglement entropy at quadratic order in $\mu^2$ 
\bea
 S_{\rm EE}(\Delta)\,=\,\frac{N}{3}\ln\Big|\frac{\beta}{\pi}\sinh\left(\frac{\pi\Delta}{\beta}\right)\Big|\,+\,\frac{5N\mu^2}{6\pi^2}\left(\frac{1}{2}\,{\cal I}_1(\Delta)
\,-\,\frac{1}{20}\,{\cal I}_2(\Delta)\right)\,+\ldots\label{eeff}
\eea
This is the result quoted in eq.\eqref{EEmain}. The outcome of integration is given in eqs.\eqref{I1final} and \eqref{I2final}, and the method of evaluation is described in appendix \ref{integrals}.

We now make a final remark regarding the U(1) current $J$ in the free fermion theory. If the spin-three current is taken to be $W\sim \sum_a:(\partial\varphi_a)^3:$ then the OPE of the stress tensor with $W$ has the form 
\be
T(z_1)W(z_2)\,\sim\,\frac{3 W(z_2)}{(z_1-z_2)^2}+\frac{W'(z_2)}{(z_1-z_2)}+
\frac{J(z_2)}{(z_1-z_2)^4}\,,
\ee
which means that $W$ is not a primary as long as $J$ is non-zero. For our calculation we have taken the two-point function of $W$ on the cylinder to be of the form expected for a primary field \eqref{2pointw}. Thus we have restricted attention to the $J=0$ sector to the order we have worked in. That we have self-consistently done so follows from the twin facts that the two-point function for $W$ not only reproduces the expected thermal correction to the free energy \eqref{thermalz} (see also \cite{Kraus:2011ds}), but also leads to an EE which yields the correct thermal entropy in the high temperature limit. It is not {\em a priori} clear that this approach will continue to be consistent at higher order in the expansion in powers of $\mu$. It would be interesting to revisit this issue to achieve a more systematic understanding.

\section{Free bosons: ${\cal W}_\infty[1]$ symmetry}
\label{freebose}
In this section we will repeat the evaluation of the entanglement entropy to order $\mu^2$ for the 
theory of $N$ complex bosons, $ X_a$ with $a = 1, \cdots N$.
The free boson case is somewhat non-trivial since the twist fields cannot be written down explicitly, in contrast to the free fermion CFT. The $N$ bosons transform under a $U(N)$ global symmetry and the theory admits a 
${\cal W}_{\infty }[\lambda]$  symmetry  with $\lambda =1$ \cite{Pope:1989ew, Bakas:1990ry}. 
The currents generating the ${\cal W}_{\infty}[1]$ algebra are built out of bilinears of the 
complex bosons and are  $U(N)$ singlets. 
As usual, the OPE of the free bosons on the plane is given by 
\begin{equation}
\bar{X}_a ( z)  X^b(w) \,\sim \,-\, \delta_{a}^b\, \ln\left| z- w\right|^2\,, 
\end{equation}
while the spin-2 and spin-3 currents  within the ${\cal W}_\infty$ tower are 
\begin{eqnarray}\label{boson-W-def}
&&T(z) \, =\,  -\, \partial X^a \partial \bar X_a,
\label{spin23boson} \\ \nonumber
\\\nonumber
&&W(z) \,=\,   \sqrt{\frac{5}{12\pi^2}}\, ( \pd ^2 \bar{X}_a \pd X^a\,  -\, \pd \bar{X}_a \pd^2 X^a)\,. 
\end{eqnarray}
Here again we have normalized the spin-3 current to reproduce the 
leading singularity as in (\ref{opew}).  For later use, we will denote the normalization of the 
spin-3 current as 
\begin{equation} \label{spin3norm}
 \tilde a = i  \sqrt{\frac{5}{12\pi^2}}.
\end{equation}
Unlike the case of the free fermion realization of the ${\cal W}_{1+\infty}$ algebra discussed in 
section \ref{freefer},  the spin-3 current  $W$ for the free boson theory 
is a primary field of the theory.

\subsection{Correlators involving bosonic twists}

As discussed in section \ref{freefer}, to evaluate the R\'{e}nyi entropies to order $\mu^2$ we need
the correlators of the twist operators with the spin-3 currents. Although the twist operator 
for the free boson field theory is not explicitly known, its  correlators with the currents 
of the theory can be written down using the knowledge of OPEs 
of the twist operator 
with these currents \cite{Dixon:1986qv}.   We will first obtain the required correlators 
on the plane and then transform them 
to the cylinder.

In this section we will omit the `flavour' index, and focus attention on a single complex boson $X$. 
The twist operator has conformal dimension $(h,\bar h)$ with
\begin{equation} \label{confdim}
 h\, =\, \bar h\,=\,   \frac{k}{2n} \left( 1- \frac{k}{n} \right) , 
 \end{equation}
 where $k$ labels the replica in Fourier space and $n$ is the total number of replicas. 
 The replica label $k$ runs form $0$ to $n-1$. 
The  two-point function of the twist operators is then given by\footnote{
Note that we have placed the anti-twist operator at $y_1$ and the twist operator at $y_2$. }
\begin{equation}
\langle\, \bar\sigma_{k, n} ( y_1, \bar y_1)\, 
\sigma_{k, n} ( y_2, \bar y_2) \rangle\, =\, 
\frac{1}{| y_1 - y_2|^{\frac{2k}{n} \left( 1- \frac{k}{n} \right) } }\,. 
\end{equation}
The spin-3 current $W$ involves a specific combination
\eqref{spin23boson} of bi-linears in derivatives of free boson fields.  Therefore, to obtain the three- and  four-point 
correlators of the form 
\begin{equation}
G_{(3)}\,=\,\langle\, W \,\bar \sigma_{k,n}\, {\sigma}_{k, n}\, \rangle \qquad \hbox{and}  \quad G_{(4)}\,=\,
\langle\, W\,W\, \bar \sigma_{k,n}\,   {\sigma}_{k, n}\, \rangle,\label{34ptboson}
\end{equation}
which will determine the order $\mu$ and $\mu^2$ corrections respectively,
we begin by first considering correlation functions involving insertions of pairs of the form  $\partial_z X \partial_w \bar X$ in the presence of twist fields:
\begin{eqnarray} \label{g-def}
&&g(z, w; y_i) \,=\, \frac{ \langle \,-\,  \partial_z X\, \partial_w \bar X\, 
\bar  \sigma_{k, n} ( y_1, \bar y_1)\,\sigma_{k, n} (y_2, \bar y_2)  
\,\rangle }{ \langle \bar\sigma_{k, n} ( y_1, \bar y_1)\, \sigma_{k, n} 
( y_2, \bar y_2) \rangle }\,, \\ \nonumber\\\nonumber
 &&h(z, w, z', w'; y_i) \,=\, 
\frac{ \langle\,\partial_z X(z)  \partial _w \bar X (w) \,
\partial_{z'} X(z')  \partial _{w'} \bar X (w') \,
\bar \sigma_{k, n} ( y_1, \bar y_1) \,\sigma_{k, n} (y_2, \bar y_2)\, \rangle }{ \langle 
\bar\sigma_{k, n} ( y_1, \bar y_1)\, \sigma_{k, n} (y_2, \bar y_2)  \rangle }\,. 
\end{eqnarray}
Once these are known, given the form of the $W$-current, we can obtain the three- and four-point functions \eqref{34ptboson} by considering derivatives of these correlators with respect 
to $z, w, z', w'$ and then taking the coincidence 
limits $z\rightarrow w \rightarrow z_1$ and $z'\rightarrow w'\rightarrow z_2$. 

Let us first consider the Green's function $g(z, w; y_i)$. From the behaviour of the OPE of the currents 
$\partial X, \partial \bar X$ with the twist and anti-twist operator given in \cite{Dixon:1986qv}  this correlator 
is required to satisfy the following properties: 
\begin{eqnarray} \label{singu}
  g( z, w\,;\, y_i) \,&\sim&\,  \frac{1}{( z- w)^2 }\, +\, \hbox{finite}, \qquad\, z \rightarrow w\, , \\ \nonumber
 & \sim&\, \frac{1}{ ( z - y_1) ^{\frac{k}{n} }} \qquad\qquad\qquad z \rightarrow y_1, \\ \nonumber
& \sim &\, \frac{1}{ ( z- y_2) ^{ 1- \frac{k}{n} } }\qquad\qquad\quad z \rightarrow y_2\,,  \\ \nonumber
& \sim &\, \frac{1}{( w - y_1)^{ 1- \frac{k}{n} } } \qquad\qquad\quad w \rightarrow y_1\, , \\ \nonumber
&\sim &\, \frac{1}{ ( w - y_2)^{ \frac{k}{n} } } \qquad\qquad\qquad  w \rightarrow y_2\,. 
\end{eqnarray}
The first of these properties results from factorization of the correlation function in the 
limit $z\rightarrow w$. The answer for $g(z,w;y_i)$ can be written down in terms of certain cut differentials which we define as
\bea \label{cutdif}
&&\omega_k(z)\, =\, ( z-y_1)^{-k/n}\, ( z-y_2)^{- ( 1- k/n)}\,, \\ \nonumber\\\nonumber
&&\omega_{n-k}(z)\, =\, ( z- y_1)^{-( 1-k/n) }\, ( z- y_2) ^{-k/n}\,. 
\eea
Then the properties listed in  eq.\eqref{singu} lead to the following result for the Green's function: 
\begin{eqnarray}
&&g(z, w\,;\,y_i) \,=\, \omega_k(z)\, \omega_{n-k}( w)\,\times\\\nonumber\\\nonumber
&&\times \left(\,   \frac{1}{(z-w)^2}\,   \left[
  \tfrac{k}{n}\,  ( z-y_1) ( w-y_2)\,  +\, \left( 1-\tfrac{k}{n} \right) ( w- y_1) ( z-y_2)   \right]
 + A( y_1, y_2, \bar y_1, \bar y_2)   \right).     \label{green-g}
\end{eqnarray}
Note that the coefficients of the first two terms, $\left(\frac{k}{n}\right)$ and $\left(1-\frac{k}{n}\right)$, ensure that the above correlator does not 
have a simple pole as $z\rightarrow w$. 
We now turn to the function $A$ which remains undetermined by the conditions (\ref{singu}). When the branch points at $y_1$ and $y_2$ are taken to approach each other, the Green's function must reduce to  $-\langle\partial_z X\,\partial_w \bar X\rangle\,=\, 1/(z-w)^2$.  This yields the requirement
\begin{equation}
 \lim_{y_1\rightarrow y_2} A =0\,. 
\end{equation}
By similar reasoning we also have the condition that 
\begin{equation}
 \lim_{y_1, y_2 \rightarrow \infty} A=0\,.  
\end{equation}
These limiting behaviours suggest that  the most natural choice is to set $A=0$.  
We will show that this choice  is  consistent with the  requirement that 
$g(z, w\,;\,y_i)$ reduces to  the three point function of the stress tensor with the twist operator
on subtracting off the $1/(z-w)^2$ divergence. 
From now on we set $A=0$, so that we obtain 
\begin{equation} \label{gexp}
g(z, w\,;\,y_i)\, = \,\omega_k(z)\, \omega_{n-k}( w)\, E(z, w)\, ,
\end{equation} 
where 
\begin{equation}\label{e-def}
E(z, w)\, =\, \frac{1}{( z-w)^2} \left( 
\tfrac{k}{n} ( z-y_1) ( w-y_2)\, +\, \left( 1-\tfrac{k}{n} \right)( w-y_1) ( z-y_2) \right). 
\end{equation}
Let us verify the consistency of the choice $A=0$. We first  evaluate the limit,
\begin{eqnarray} \label{streslim}
 \lim_{z\rightarrow w} \left( g(z,w) - \frac{1}{(z-w)^2} \right) 
\,=\,  \frac{k}{2n}\,\left ( 1- \frac{k}{n} \right) \left( 
\frac{1}{ w - y_1}  - \frac{ 1}{ w- y_2} \right)^2  . 
\end{eqnarray}
Recall that the standard OPE for the current $\partial X$  with $\partial \bar X$ and  in particular, its finite part yields the stress tensor:
\begin{equation}
\partial X(z) \partial \bar X(w)\, \sim\, \frac{1}{(z-w)^2}\, +\, T(w) \,+ \,\cdots
\end{equation}
Therefore the limit in eq.(\ref{streslim}) allows us to deduce that the three-point function involving the stress tensor and the twist fields is,
\begin{equation}
 \langle \,T(w)\, \bar\sigma_{k, n} ( y_1)\, \sigma_{k, n} ( y_2)\, \rangle\, 
=\,\frac{k}{2n} \left( 1- \frac{k}{n} \right)\, 
\frac{1}{( y_1 -y_2) ^{2 \Delta_k - 2} ( w -y_1)^2 ( w- y_2) ^2 }\,. 
\end{equation}
This is precisely the form expected on the basis of conformal invariance. The normalization is also consistent with the 
OPE  of the stress tensor with the twist operator which is primary with conformal dimension given in eq.\eqref{confdim}. 

We can now proceed to construct the Green's function $h(z, w, z', w', y_i)$. The function which satisfies all the
necessary properties of the correlator involving two insertions of $\partial_z X \partial_w \bar X$ in the presence of twist fields is given by 
\bea \label{hexp}
 &&h(z, w, z', w';y_i)\,=\\\nonumber
 &&\qquad\omega_k(z)\, \omega_k(z') \,\omega_{n-k}(w)\, \omega_{n-k}(w')\, 
\left( E(z, w)\, E(z', w')\, +\, E( z', w)\, E( z, w')  \right)\,, 
\eea
where $\omega_k$ and $E$ are defined in eqs.\eqref{cutdif} and \eqref{e-def}.
This function exhibits the correct singular behaviour when the positions of the current insertions approach 
that of the twist fields, as discussed earlier. 
It also has the required factorization properties  in any of the limits,
$z\rightarrow w$,  $z'\rightarrow w'$,  $z\rightarrow w'$ and $z'\rightarrow w$. 
The anti-holomorphic counterparts of the Green's functions we found above are, respectively, 
\bea \label{g-defa}
\bar g(\bar z, \bar w; \bar y_i) \,&=&\,
  \frac{ \langle - \, \partial_{\bar z} X(\bar z) \, \partial_{\bar w} \bar X (\bar w)\,  
 \bar \sigma_{k, n}  ( y_1, \bar y_1)\,\sigma_{k, n} (y_2, \bar y_2)\, \rangle }{ \langle\, \bar\sigma_{k, n} ( y_1, \bar y_1)\, \sigma_{k, n} ( y_2, \bar y_2)\, \rangle }, \\\nonumber\\\nonumber
\,&=&\,  \omega_{n-k}(\bar z)\, \omega_{k}( \bar w)\, E(\bar z, \bar w)\,,
\eea
and
\bea
&&  \bar h(\bar z, \bar w,  \bar z', \bar w'; \bar y_i)\,=\, 
 \frac{ \langle(  - \partial_{\bar z} X( \bar z)  \partial _{\bar w} \bar X ( \bar w) )
 (  -\partial_{\bar z'} X(z')  \partial _{\bar w'} \bar X (\bar w') )
 \bar \sigma_{k, n} ( y_1, \bar y_1) \sigma_{k, n} (y_2, \bar y_2) \rangle }{ \langle 
 \bar \sigma_{k, n} ( y_1, \bar y_1) \sigma_{k, n} (y_2, \bar y_2)  \rangle },\nonumber \\\nonumber\\
 &&=\,
 \omega_{n-k}(\bar z)\, \omega_{n-k} (\bar z') \,\omega_{k}(\bar w)\, \omega_{k}(\bar w') 
\left( E(\bar z, \bar w) E(\bar z', \bar w') + E( \bar z',  \bar w) E( \bar z,  \bar w')  \right). 
\eea
We note that these anti-holomorphic counterparts can be obtained from the holomorphic ones 
by replacing the co-ordinates with their complex conjugates and the making the replacement $k\rightarrow n-k$.  This ensures the required singularity structure in the limit that the current insertions approach the twist operators.

Finally  it is important to note that when 
there are $N$ free bosons in the theory, the twist operator is constructed by the 
product of the twists for each boson. Therefore we have
\begin{equation} \label{ptwist}
 \sigma_{k, n} (y, \bar y) = \prod_{a =1}^N \sigma_{k, n }^{(a) } (y, \bar y)\,, 
\end{equation}
where the superscript $(a)$ labels the corresponding boson on which the twist acts. 
Given the Green's functions  $g(z, w;y_i)$, 
$h(z, w, z', w'; y_i) $  and their complex conjugates,  
 the correlators involving the spin-3 currents which yield  the R\'{e}nyi or entanglement
 entropies can now be found by appropriate differentiation of $g$ and $h$ and then taking coincidence limits for some of the arguments of these functions.

%\subsection{Calculating $\langle \sigma \bar \sigma W W  \rangle$}

\subsection*{The three-point function}
Using all the ingredients discussed above we can now determine the  3-point function involving the twists with a single insertion of the spin-3 current, given by 
\begin{align}
G_{(3)}\,=\,\langle\, \bar \sigma_{k, n}  (y_1, \bar y_1)\,  \sigma_{k, n} (y_2, \bar y_2)\, W (z)\, \rangle.  \nn 
\end{align}
This 3-point function will determine the contribution to R\'{e}nyi  entropies at first order in $\mu$. 
Using the definitions of $W$ and $g$ from eqs.\eqref{boson-W-def} and \eqref{g-def} respectively, we find
\begin{align} \label{3ptfnw}
\frac{\langle\, \bar\sigma_{k, n}
 (y_1, \bar y_1)\,  \sigma_{k, n} (y_2, \bar y_2)\, W (z) \,\rangle }
 {\langle \,\bar\sigma_{k, n} (y_1, \bar y_1)\,  \sigma_{k, n} (y_2,\bar y_2)\, \rangle} \,&=\, -\,\tilde a \lim_{z \rightarrow w}\, \left(\pd_z g(z,w)\, -\, \pd_w g(z,w) \right) \nn \\
&\,=\,-\tilde a\,N\,   \frac{k\, (k-n)\, (2 k-n)\, (y_1-y_2)^3}{3 n^3\, (z-y_1)^3\, (z-y_2)^3}\,. 
\end{align}
The factor of $N$ in the second line results from the sum over $N$ species of free bosons
and the fact the twist field (\ref{ptwist}) is a product of the twist fields for each of the $N$ bosons.
The normalization $\tilde a$ is defined in (\ref{spin3norm}).  We note that each of the limits: $\lim_{z\rightarrow w} \partial_z g(z, w)$ and 
$\lim_{z\rightarrow w } \partial_w g(z, w)$, contains singularities. These singularities precisely 
cancel each other in eq.(\ref{3ptfnw}) leading to a finite result. 

Since all the fields in the three-point function are primaries we can easily transform the result 
(\ref{3ptfnw}) on the plane to the 
cylinder using the conformal  transformation given in 
eq.(\ref{transcy}). Thus the three-point correlator on ${\mathbb R}\times S^1_\beta$ for the free boson theory is
\begin{align} \label{3pt1}
\frac{\langle \bar\sigma_{k, n}
 (y_1, \bar y_1)\,  \sigma_{k, n} (y_2, \bar y_2)\, W (z) \rangle }
 {\langle \bar\sigma_{k, n} (y_1, \bar y_1)\,  \sigma_{k, n} (y_2,\bar y_2) \rangle} 
\,=\, -\tilde a\,N   \frac{k\, (k-n)\, (2 k-n) }{3\, n^3 }\, G(z)^3, 
\end{align}
where now the corresponding coordinates are on the cylinder and the function $G(z)$ is defined in \eqref{HGdef}. 
Repeating the same calculations for the anti-holomorphic insertion of the current we find
\begin{align} \label{3pt2}
\frac{\langle \bar\sigma_{k, n}
 (y_1, \bar y_1)\,  \sigma_{k, n} (y_2, \bar y_2)\, \bar W (\bar z) \rangle }
 {\langle \bar\sigma_{k, n} (y_1, \bar y_1)\,  \sigma_{k, n} (y_2,\bar y_2) \rangle}  
&=\, \tilde a\, N\,   \frac{k\, (k-n)\, (2 k-n)\, }{3 n^3 }\, G(\bar z)^3. 
\end{align}
The anti-holmorphic three-point function can be obtained from the holomorphic one
by replacing the holomporphic co-ordinates by their anti-holomorphic counterparts
and by replacing $k\rightarrow n-k$. 

\subsection*{The four-point functions}

We now turn to the 4-point function involving the insertion of two spin-3 currents. We first consider the 4-point function which involves only the holomorphic  insertion of the 
spin-3 currents.  We can decompose the contributions to this correlator into two distinct pieces
as shown below
\begin{align}
& {\langle\, \bar \sigma_{k, n} (y_1,\bar y_1)\,  \sigma_{k, n}  (y_2,\bar y_2)\, W(z_1)\, W(z_2)\,  \rangle}\,=\, \\& \tilde a^2 \sum_{a=1}^{N} \left\langle \bar\sigma_{k, n}    \sigma_{k, n} ( \pd ^2 \bar{X}_a \pd X^a  - \pd \bar{X}_a \pd^2 X^a   )(z_1) ( \pd ^2 \bar{X}_a \pd X^a  - \pd \bar{X}_a \pd^2 X^a   )(z_2)  \right\rangle\,+\, \nn \\\nonumber
& \tilde a^2 \sum_{a,b=1 \ ;\ a\neq b}^{N}\left\langle \bar \sigma_{k, n}
\sigma_{k, n} \left( \pd ^2 \bar{X}_a \pd X^a  - \pd \bar{X}_a \pd^2 X^a   \right)(z_1)\, \left( \pd ^2 \bar{X}_b \pd X^b  - \pd \bar{X}_b \pd^2 X^b \right)(z_2)  \right\rangle\,,
\end{align}
where in the second and third lines the positions of the twist fields have been omitted for clarity of presentation.
In the expression above, the first term is a sum over the $N$ species of the type $\sum_a W_{(a)} W_{(a)}$, where $W_{(a)}$ can be viewed as the spin-3 current associated to the individual free boson $X^a$. The second sum, on the other hand, is of the type $\sum_{a\neq b} W_{(a)} W_{(b)}$, and there are $N(N-1)$ terms in the sum.
%result from the same boson bi-linears from  both the spin-3 current, while
%the second set of terms result from non-identical boson bi-linears form the spin-3 currents. 
%The first set of terms are $N$ in number, while the second set of terms are 
%$N(N-1)$ in number. 

It is clear that the correlator involving the first set of terms $\langle\bar \sigma \sigma \sum_a W_{(a)} W_{(a)}\rangle$ should be completely determined by the Green's function $h$
in eq.\eqref{g-def} for the single free boson.  Since bosons 
with different species index have a trivial OPE, the second set of terms $\langle\bar \sigma \sigma \sum_{a\neq b} W_{(a)} W_{(b)}\rangle$ are determined by 
the Green's function $g$ in eq.\eqref{g-def}. 
Therefore we obtain  
\begin{align}\label{boson-4-pt-01}
&\frac{\langle\, \bar\sigma_{k,n}(y_1,\by_1)\,  \sigma_{k,n} (y_2,\by_2)\, W(z_1)\, W(z_2)\,  \rangle}{\langle\, \bar \sigma (y_1, \by_1)\, \sigma (y_2,\by_2) \rangle}  \\&= \lim _{z_1' \rightarrow z_1 \ ; \ z_2' \rightarrow z_2 } \Bigg[ \tilde a^2  N  \left( \pd_{z_2}\pd_{z_2'}h \left(z_1,z_2,z_1',z_2'\right)\,-\,\pd_{z_1}\pd_{z_2'}h \left(z_1,z_2,z_1',z_2'\right)\nn \right.\\
&\left. \hspace{4cm}-\,\pd_{z_2}\pd_{z_1'}h \left(z_1,z_2,z_1',z_2'\right)   \,+\,\pd_{z_1}\pd_{z_1'}h \left(z_1,z_2,z_1',z_2'\right) \right) \nn \\\nonumber\\
&\quad +\, \tilde a^2  N (N-1)\,  \Big( \pd_{z_2}g \left(z_1,z_2\right)\pd_{z_2'}g \left(z_1',z_2'\right)-\pd_{z_1}g \left(z_1,z_2\right)\pd_{z_2'}g \left(z_1',z_2'\right)\nn \\\nonumber
&\hspace{4cm}-\pd_{z_2}g \left(z_1,z_2\right)\pd_{z_1'}g \left(z_1',z_2'\right)+\pd_{z_1}g \left(z_1,z_2\right)\pd_{z_1'}g \left(z_1',z_2'\right)\Big) \Bigg] 
\end{align}
Using our explicit expressions for the  Green's functions given in eqs.\eqref{gexp} and \eqref{hexp} the above quantities in parenthesis can be calculated. Taking the limits $z_1' \rightarrow z_1$ and $ z_2'\rightarrow z_2$, we find
\begin{align}\label{boson-cross}
& \pd_{z_2}g \left(z_1,z_2\right)\pd_{z_2'}g \left(z_1',z_2'\right)-\pd_{z_1}g \left(z_1,z_2\right)\pd_{z_2'}g \left(z_1',z_2'\right)-\pd_{z_2}g \left(z_1,z_2\right)\pd_{z_1'}g \left(z_1',z_2'\right) \nn \\\nonumber
& \hspace{3.5in}+\pd_{z_1}g \left(z_1,z_2\right)\pd_{z_1'}g \left(z_1',z_2'\right)\nn \\\nonumber\\
&\xrightarrow{z_1' \rightarrow z_1 \ ; \ z_2' \rightarrow z_2 } \ \  \frac{k^2 \left(y_1-y_2\right){}^6 (k-n)^2 (2 k-n)^2}{9 n^6 \left(z_1-y_1\right){}^3 \left(z_1-y_2\right){}^3 \left(z_2-y_1\right){}^3 \left(z_2-y_2\right){}^3}
\end{align}
and 
\begin{align}
&\pd_{w}\pd_{w'}h \left(z_1,z_2,z_1',z_2'\right)-\pd_{z_1}\pd_{z_2'}h \left(z_1,z_2,z_1',z_2'\right)-\pd_{z_1}\pd_{z_1'}h \left(z_1,z_2,z_1',z_2'\right) \\\nonumber
 &\hspace{3.5in}+\,\pd_{z_1}\pd_{z_1'}h \left(z_1,z_2,z_1',z_2'\right)\\\nonumber \\\nonumber     
&\xrightarrow{z_1' \rightarrow z_1 \ ; \ z_2' \rightarrow z_2 } \ \  
\frac{4}{\left(z_1-z_2\right)^6}\,+\,
\frac{k^2 \left(y_1-y_2\right){}^6 (k-n)^2 (2 k-n)^2}{9 n^6 \left(z_1-y_1\right){}^3 \left(z_1-y_2\right){}^3 \left(z_2-y_1\right){}^3 \left(z_2-y_2\right){}^3}\nn \\ 
\nonumber\\\nonumber 
& \hspace{1in}
-\frac{k \left(y_2-y_1\right){}^4 (k-n) (2 k-n)^2}{n^4 \left(z_1-y_1\right){}^2 \left(y_2-z_1\right){}^2 \left(z_1-z_2\right)^2 \left(y_2-z_2\right){}^2 \left(z_2-y_1\right){}^2}\nn \\\nonumber\\\nonumber  & 
\hspace{1in}
-\frac{6 k \left(y_2-y_1\right){}^2 (k-n)}{n^2 \left(z_1-y_1\right) \left(y_2-z_1\right) \left(z_1-z_2\right)^4 \left(y_2-z_2\right) \left(z_2-y_1\right)}\,.\nn 
\end{align}
Putting together these results and after performing the conformal transformation to the cylinder ${\mathbb R}\times S^1_{\beta}$ we arrive at the following expression for the 4-point function\footnote{Note that here $$W(z_1)W(z_2)=\sum_{i}\sum_{j} W_i(z_1)W_j(z_2)$$ $i,j$ being the Fourier indices and running from $0$ to $n-1$. The RHS of \eqref{4pt1} is the expression after these $i$ and $j$ sums are done.} (see eq.\eqref{HGdef} for the definition of the functions $H$ and $G$):
\begin{align} \label{4pt1}
 &\frac{\langle\, \bar\sigma_{k,n}(y_1,\by_1)\,  \sigma_{k,n} (y_2,\by_2)\, W(z_1)\,   W(z_2)\,\rangle }{  \langle\,
  \bar \sigma_{k,n}(y_1,\by_1) \,\sigma_{k,n} (y_2,\by_2) \,\rangle}\,=\, \tilde a^2 N \left[\,4 H(z_1-z_2)^6 \,\right.\nonumber \\\nonumber\\\nonumber
&\left. -\, \kappa\left(\kappa-1\right)\left(2\kappa-1\right)^2G(z_1) ^2G(z_2)^2H(z_1-z_2)^2 \,-\,6\kappa \left(\kappa-1\right)G(z_1)\, 
G(z_2)\, H(z_1-z_2)^4\right]\nn \\\nonumber\\
 & \qquad\qquad \text{with } \kappa=k/n\,.  
%\label{4pt2}
\end{align} 
The insertions of  purely anti-holomorphic spin-3 currents are given by exactly the same formula with $z_{1,2}\to\bar z_{1,2}$.

Finally, there is also a cross term at order $\mu^2$ which involves one insertion each of 
the holomorphic spin-3 current $W$ and its anti-holomorphic counterpart $\overline W$. 
Since the $W\overline W$ OPE is trivial, the four-point function $\langle\sigma\bar\sigma W\overline W\rangle$ factorizes and can be 
written in terms of the product of the three point functions in eqs.\eqref{3pt1} and \eqref{3pt2}. Each of these three-point functions independently vanishes upon performing the $k$-sums. This is due to the fact (as discussed below \eqref{onept}) that the one-point functions of the spin-three primary current vanish on the plane and therefore on the $n$-sheeted Riemann surface.
%\bea
%&&\frac{\prod_{k}\langle\, \bar\sigma_{k,n}(y_1,\by_1)\,   \sigma_{k,n} %(y_2,\by_2)\, W(z_1)\, \bar W(\bz_2)\,  \rangle }{ \prod_k \langle \bar%\sigma_{k,n}(y_1,\by_1)\,  \sigma_{k,n} (y_2,\by_2) \rangle}\\\nonumber\\\nn
%&&\hspace{1.2in}\,=\,-\,\frac{\tilde a^2\, N^2}{9}\,\prod_{k}{\kappa}\left%(1-\kappa\right) \left(1-2\kappa\right)\prod_{k'} {\kappa'}\left(1-%\kappa'\right) \left(1-2\kappa'\right)   
%G(z_1) ^3 G(\bz_2)^3\,\,;\qquad\kappa=k/n\,.\label{boson-4-pt-cross}
%\eea

\subsection{R\'{e}nyi and Entanglement entropies at ${\cal O}(\mu^2)$ }

We now have all the ingredients necessary to evaluate the R\'{e}nyi and entanglement 
entropies in the free boson theory to order $\mu^2$.   
The 3-point functions (\ref{3pt1}) , (\ref{3pt2})  
 involving a single insertion of the spin-3 current potentially contribute at ${\cal O} (\mu)$ while
 the 4-point function (\ref{4pt1}) and its anti-holomorphic counterpart
 contribute at ${\cal O}(\mu^2) $. 
 We will now discuss each of these in some more detail. 

\subsection*{First order correction in $\mu$}

We will set $y_1=0$ and $y_1 = \Delta $ where $\Delta$ is real. 
This ensures that we are evaluating the entanglement entropy associated to a spatial 
interval. Summing up the contributions of the holomorphic and anti-holomorphic insertions of the 
$W$ current we obtain 
\bea
&&-\mu\int d^2 z\,\langle\sigma_{k,n}(y_1,\bar y_1) \,\bar\sigma_{k,n}(y_2,\bar y_2) 
\left(W(z)\,+\,\overline W(\bar z)\right)\rangle
\\\nonumber
&&=\, \frac{i\,N}{3}\sqrt{\frac{5}{12\pi^2}} 
 \kappa(\kappa-1) ( 2\kappa-1)
{{\Big|\frac{\beta}{\pi}\sinh\left(\frac{\pi\Delta}{\beta}\right)\Big|^{-2\kappa^2N}}} \int d^2 z
\left[G(z)^3\,-\,G(\bar z)^3
\right]\,=\,0\,,
\eea
where $\kappa=k/n$.
Here we have substituted in the 3-point functions derived in eqs.\eqref{3pt1} and \eqref{3pt2}. We have chosen   equal chemical potentials 
for both holomorphic and anti-holomorphic sectors ($\mu=-\bar \mu$) which results in the cancellation of the 
linear term in $\mu$, in the same fashion as in the free fermion theory.  As explained in the discussion below eq.\eqref{ff1pt} this is a consequence of the general fact that one point functions of the spin-three current must vanish on the plane and therefore, by the tensor transformation law for primaries, on the $n$-sheeted Riemann surface as well. The result holds also for  unequal chemical potentials as for the free fermion
theory, and it follows that the three point function 
involving the twists and a single insertion of the spin 3 current vanishes. 
Thus  the contribution at first order in $\mu$ vanishes even for $\mu \neq -\bar \mu$. 

%performing the 
%sum over $k$,  which runs from $0$ to $ n-1$, ensures that the linear term proportional to $(\mu +\bar \mu)$ in the R\'{e}nyi entropy still  vanishes. It is interesting to note that apart from the normalization constant, the holomorphic and anti-holomorphic spin-3 charges (obtained by integrating the currents) in the presence of the twist fields are  identical to those in the free fermion theory \eqref{onept}.

\subsection*{Second order correction in $\mu$}

For the second order contribution we obtain 
\bea \label{bosoncon}
&&\tfrac{1}{2}\int d^2z_1\int d^2z_2
\left\langle\, \sigma_{k,n} (y_1,\bar y_1)\, 
\left[W(z_1)+\overline W(\bar z_1)\right]\,\left[W(z_2)+\overline W(\bar z_2)\right]\,\overline{\sigma}_{k,n} (y_2,\bar y_2) \,  \right\rangle_{\rm CFT}
\nonumber\\\nonumber\\
&&=- \frac{5N}{12\pi^2} \,\Big|\frac{\beta}{\pi}\sinh\left(\frac{\pi\Delta}{\beta}\right)\Big|^{-2N\frac{k}{n}
\left( 1- \frac{k}{n} \right) }\int d^2z_1\int d^2z_2\left( \, 4\,H^6(z_1-z_2)\,+
\right.\\\nonumber\\\nonumber
&&\left.- \frac{6k(k-n)}{n^2} 
H^4(z_1-z_2)G(z_1)G(z_2)-\frac{k(k-n)(2k-n)^2}{n^4} \,H^2(z_1-z_2)G^2(z_1)G^2(z_2)\right).
\eea
The terms proportional to $N^2$ in the individual four-point functions \eqref{4pt1} disappear due to a cancellation once the $k$ sum is carefully performed.

The structure of eq.\eqref{bosoncon} is similar to the corresponding contribution \eqref{4ptff} in the free fermion CFT, although it is clear that $k$-dependence of the coefficients is different. 

A first check of our calculation so far is provided by the first term in (\ref{bosoncon}) which is proportional to $H^6(z_1-z_2)$. After integration, it yields an extensive answer which, upon including the overall normalization, matches the expected ${\cal O}(\mu^2)$ correction to the thermal partition function for a theory with central charge $c=2N$. The free fermion case also exhibits the very same feature.
This extensive contribution is cancelled when we divide by the normalization $Z^n$ in the 
R\'{e}nyi entropy (\ref{renyi}).  The R\'{e}nyi entropy  can  be written as a 
sum over $k$ which labels the replicas as 
\begin{eqnarray}
S^{\,(n)} = \frac{1}{1-n} \sum_{k =0}^{n-1}  \ln\left( \frac{Z_k}{Z} \right)\,. 
\end{eqnarray}
Working to quadratic order in $\mu$,
\bea
&&
Z_k\,=\,\Big|\tfrac{\beta}{\pi}\sinh\left(\tfrac{\pi\Delta}{\beta}\right)
\Big|^{-2N\frac{k}{n}\left( 1- \frac{k}{n}\right)}\times\\\nonumber\\\nonumber
&&\hspace{0.5in}\left[1\,+
\, 2N\,\frac{8\pi^3}{9}\frac{\mu^2L}{\beta^3}\,+\,
\frac{5N\mu^2}{12\pi^2}
\left(
\tfrac{6k(k-n)}{n^2}\,{\cal I}_1(\Delta)\,+
\,\tfrac{k(k-n)(2k-n)^2}{n^4}\,{\cal I}_2(\Delta)
\right)+\ldots\right]\,,
\eea
where ${\cal I}_1$ and ${\cal I}_2$ are the integrals defined in eqs.\eqref{I1} and \eqref{I2}.
The sums over $k$ are given by the formulae
 \begin{align}
& \sum _{k=0}^{n-1} \frac{k (k-n)}{n^2}\, =\,-\frac{(n-1) (n+1)}{6 n}\,,\\
&\sum _{k=0}^{n-1} \frac{k (k-n) (2 k-n)^2}{n^4}\, =\, -\frac{(n-2) (n-1) (n+1) (n+2)}{30 n^3}\,,
 \end{align}
which, upon substitution into the R\'enyi entropy and taking the limit $n\rightarrow 1$ yield
the entanglement entropy at order $\mu^2$ in the free boson CFT:
\be
 S_{\rm EE}(\Delta)\,=\,\frac{2N}{3}\ln\Big|\frac{\beta}{\pi}\sinh\left(\frac{\pi\Delta}{\beta}\right)\Big|\,+\,\frac{10 N\mu^2}{6\pi^2}
\left(\frac{1}{2}\,{\cal I}_1(\Delta)\,-\,\frac{1}{20}
\,{\cal I}_2(\Delta)\right)\,+\ldots
\ee
Identifying the  central charge $c=2N$, this agrees precisely with our main result given in equation 
\eqref{EEmain}. The integrals ${\cal I}_1, {\cal I}_2$ are given in equations
eqs.\eqref{I1final} and \eqref{I2final}. We have therefore established by explicit computation that the order $\mu^2$ correction to the entanglement entropy is identical for the two ${\cal W}_\infty[\lambda]$ CFTs with $\lambda=0$ and $\lambda=1$.

It is interesting to note that while the entanglement entropies at order $\mu^2$ for the free boson and free fermion CFTs agree with each other, the R\'enyi entropies $S^{\,(n)}$ are in fact different. The expression for the R\'enyi entropy in the free boson theory is  
\begin{align}
S^{\,(n)}(\Delta)&\,=\,\frac{c(n+1)}{6n}\,\log\Big|\frac{\pi}{\beta}\sinh\left(\frac{\pi\Delta}{\beta}\right)\Big|\, \nn \\ 
&+\,\frac{c\, 5\mu^2}{6\pi^2}\,\left[\,\frac{(n+1)}{4n}\,{\cal I}_1\left(\Delta\right)\,+\,\frac{(n-2)(n+1)(n+2)}{120 n^3}\,{\cal I}_2\left(\Delta\right) \right]\,+\ldots,
\end{align}
which should be contrasted with \eqref{renyiff} for free fermions. In particular, this means that the correlator of two $W$-currents in the $n$-replicated theory is certainly not universal. However, the surprising agreement of  entanglement entropies between the two CFTs with $\lambda=0$ and $\lambda=1$  suggests that the $n\to 1$ limit (of the order $\mu^2$ correction) may be universal and $\lambda$-independent.

\section{Holographic entanglement entropy}
\label{holEE}

Recently, two proposals have been advanced for calculating the holographic entanglement entropy of higher spin theories of gravity on AdS$_3$ related holographically to ${\cal W}$-algebra CFTs \cite{deBoer:2013vca, Ammon:2013hba}. Both proposals are based on Wilson lines in the bulk Chern-Simons formulation of higher spin theory. As already pointed out earlier, our CFT calculations are based on holomorphic (plus anti-holomorphic) deformations and the resulting picture is most naturally compatible with the so-called holomorphic formulation of the bulk holographic thermodynamics. Although the holomorphic formulation is not natural from a gravity perspective, nevertheless a proposal for higher spin holographic entanglement entropy compatible with holomorphic thermodynamics was put forward by de Boer and Jottar in \cite{deBoer:2013vca}. We will solely focus attention on their proposal 
but it is  also possible  to adapt the proposal of \cite{Ammon:2013hba} to the holomorphic picture.

Within the framework of ${\rm SL}(N,{\mathbb R})\times {\rm SL}(N,{\mathbb R})$ Cherns-Simons theory which describes a finite tower of higher spin fields (including gravity) with spins $s\leq N$ on AdS$_3$, classical solutions are simply provided by flat ${\rm sl}(N)$ connections $(A, \bar A)$,
\be
A\,=\,A_z\,dz\,+\,A_{\bar z}\, d\bar z\,+\,A_\rho d\rho\,,\qquad\qquad
\bar A\,=\,\bar A_z\,dz\,+\,\bar A_{\bar z} \,d\bar z\,+\,\bar A_\rho d\rho\,.
\ee
Here $\rho$ is the radial coordinate on AdS$_3$ while $(z,\bar z)$ are the coordinates in the boundary CFT. The $\rho$-dependence of the solutions is usually fixed in a simple way by utilizing the gauge freedom. Although the most general solutions for the flat connections can depend on $(z,\bar z)$ we will only be interested in backgrounds that have no such dependence.

According to the prescription of \cite{deBoer:2013vca}, the holomorphically factorized entanglement entropy for the spatial interval $\Delta$ is determined by the functional $W^{\rm holo}_{\cal R}$
(not to be confused with the spin-3 current)
\begin{align} 
S^{\rm holo}_{\textrm{EE}}(\Delta)=\dfrac{c}{24}\,\textrm{ln}\,W^{\rm holo}_{\cal R}\label{hol-dB-J}
\end{align}
with
\begin{align}
W^{\rm holo}_{\cal R}(P,Q)=\lim_{\rho_0\rightarrow\infty}\,\textrm{tr}_{\cal R}\Bigg[{\cal P}\,\textrm{exp}\int\limits_{P}^{Q}\bar{A}_{\bar{z}}\,d\bar{z}\,\,\,{\cal P}\,\textrm{exp}\int\limits_{P}^{Q}A_{z}\,dz\Bigg]\label{funct-W}
\end{align}
The formula we use differs slightly from \cite{deBoer:2013vca} as we have consistently performed the Euclidean continuation $x_+\to z$ and 
$x_-\to -\bar z$.
Here $P$ and $Q$ represent the endpoints of the entangling interval on the conformal boundary ($\rho_0\to\infty$) of AdS$_3$, and the limit instructs us to consider only the leading contribution in an expansion in powers of the (UV) cutoff $\Lambda=\textrm{exp}(\rho_0)$.

Most important of all is the choice of representation ${\cal R}$ in which to compute the Wilson line functional. The choice of representation is dictated both by the choice of embedding of ${\rm sl}(2)$ (gravity) in ${\rm sl}(N)$ and crucially, the requirement that for large interval lengths $\Delta \gg \beta$, the entanglement entropy should be extensive in $\Delta$ and coincide with the (holomorphic) thermal entropy.

For the principal embedding of ${\rm sl}(2)$ in ${\rm sl}(N)$, the Wilson line is to be evaluated in the representation with highest weight given by the Weyl vector, and ${\rm dim}({\cal R})\,=\, 2^{N(N-1)/2}$. For $N=3$ the required representation is the adjoint which has dimension 8.

At finite temperature, boundary CFT states carrying higher spin charges are (in the limit of large central charge) dual to black hole backgrounds with chemical potentials for the corresponding charges. Such classical saddles are solutions of the flatness condition for the bulk Chern-Simons theory, satisfying certain smoothness conditions which require triviality of the holonomy of the gauge connections around the (Euclidean) thermal circle \cite{Gutperle:2011kf}.  

In the radial gauge, the connections describing a static spin--three black hole in ${\rm SL}(3)\times {\rm SL}(3)$ Chern-Simons theory read
\begin{align}
A=&\left(\textrm{e}^{\rho}\,L_{+1}-\dfrac{2\pi\mathcal{L}}{k}\,\textrm{e}^{-\rho}\,L_{-1}-\dfrac{\pi\mathcal{W}}{2k}\,\textrm{e}^{-2\rho}\,W_{-2}\right) dz\label{a}\\
&-\mu\left(\textrm{e}^{2\rho}\,W_{+2}-\dfrac{4\pi\mathcal{L}}{k}\,W_{0}+\dfrac{4\pi^{2}\mathcal{L}^{2}}{k^{2}}\,\textrm{e}^{-2\rho}\,W_{-2}+\dfrac{4\pi\mathcal{W}}{k}\,\textrm{e}^{-\rho}\,L_{-1}\right)d\bar{z}+L_{0} d\rho\nonumber
\end{align}
\begin{align}
\bar{A}=&+\mu\left(\textrm{e}^{2\rho}\,W_{-2}-\dfrac{4\pi\mathcal{L}}{k}\,W_{0}+\dfrac{4\pi^{2}\mathcal{L}^{2}}{k^{2}}\,\textrm{e}^{-2\rho}\,W_{+2}-\dfrac{4\pi\mathcal{W}}{k}\,\textrm{e}^{-\rho}\,L_{+1}\right)dz\label{a-bar}\\
&-\left(\textrm{e}^{\rho}\,L_{-1}-\dfrac{2\pi\mathcal{L}}{k}\,\textrm{e}^{-\rho}\,L_{+1}+\dfrac{\pi\mathcal{W}}{2k}\,\textrm{e}^{-2\rho}\,W_{+2}\right)d\bar{z}-L_{0}d\rho\,.\nonumber
\end{align}
Here $(L_0,L_{\pm1},W_0,W_{\pm1}, W_{\pm2})$ are generators of ${\rm SL}(3,{\mathbb R})$ with $(L_0, L_{\pm1})$ constituting a three dimensional representation of ${\rm sl}(2)$. The subscripts on the generators denote their weights with respect to $L_0$. The Chern-Simons level, denoted by $k$, needs to be large so that the semiclassical picture applies. This in turn implies that the central charge $c$ of the boundary CFT must be large, since we have the relation (for the principal embedding)
\be
c\,=\,6k\,.
\ee
The  Chern-Simons connections $(A, \bar A)$ can also be re-expressed in terms of $\rho$-independent flat connections $(a,\bar a)$, where the dependence on the radial coordinate is obtained via a gauge transformation generated by $L_0$
\bea
A \,=\, b^{-1}\,a b + b^{-1} db\,,\qquad
\bar A \,=\, b\,\bar a\, b^{-1} + b\,db^{-1}\,,\qquad
b= e^{L_0\rho}\,.
\eea
The chemical potential $\mu$ for spin-three charge in the boundary CFT leads to a non-zero spin-three charge $2\pi{\cal W}$, while the black hole mass is given by ${2\pi{\cal L} }$. The dependence of the charge and mass on $\mu$ and the temperature $T$ follows entirely from the holonomy/smoothness condition:
\be
{\rm Hol}_t(A)\,=\,{\cal P}\exp\left(\oint A_t\right)\,=\,{\bf 1}\,,
\ee
and similarly for $\bar A$. This condition is equivalent to requiring that the eigenvalues of the constant matrix $\beta a_t$ be given by $(2\pi i, 0, -2\pi i)$, so that
\be
{\rm det}(\beta a_t)\,=\,0\,,\qquad {\rm tr}(\beta^2 a_t^2)\,=\,-8\pi^2\,,\qquad a_t\,=\,i(a_z + a_{\bar z})\,.
\ee
It turns out that the connections (\ref{a}) and (\ref{a-bar}), when coupled with the holonomy conditions which are two algebraic conditions on the charges $({\cal L},{\cal W})$, lead to four branches\footnote{An investigation of the status of multiple branches in the canonical formalism has been carried out in \cite{Chowdhury:2013roa}. For a detailed study of multiple branches and ensuing phase structure in different higher spin theories we refer the reader to \cite{Chen:2012ba}.}
of black-hole solutions \cite{D-F-K}. In what follows, we will focus on the BTZ branch, the only solution to the holonomy constraints which reduces to the BTZ black hole in the limit $\mu\rightarrow 0$. It is this branch that is smoothly connected to the undeformed boundary CFT at finite temperature.
Crucially, for the BTZ branch, the holonomy conditions impose the following relation between the chemical potential $\mu$ and the mass $\mathcal{L}$ of the black hole,
\begin{align}
\mu=\dfrac{3\sqrt{C}}{4(2C-3)}\sqrt{\dfrac{k}{2\pi\mathcal{L}}}\,,\label{mu-BTZ-branch}
\end{align}
where $C$ characterizes the dimensionless parameter $\mu T$ as
\begin{align}
\mu T=\dfrac{3}{4\pi}\dfrac{(C-3)\sqrt{4C-3}}{(3-2C)^{2}}\,.\label{C-def}
\end{align}
These are all the ingredients required to calculate the holographic entanglement entropy using the functional (\ref{funct-W}) in the adjoint representation of ${\rm sl}(3)$.
It was already pointed out in \cite{deBoer:2013vca} that in this case, in terms of the eigenvalues $(0,\pm\lambda_1,\pm\lambda_2,\pm\lambda_3)$ of the matrix $a_z$ in the adjoint representation, the Wilson line functional can be expressed as,
\begin{align}
W^{\rm holo}_{\rm adj}=\Bigg[\dfrac{8\Lambda^4}{\lambda_{1}\lambda_{2}\lambda_{3}}\Bigg]^{2}\Bigg[\dfrac{\lambda_{1}^{2}-\lambda_{2}\lambda_{3}}{\lambda_{1}\lambda_{2}\lambda_{3}}+\dfrac{\textrm{cosh}(\lambda_{1}\Delta)}{\lambda_{1}}-\dfrac{\textrm{cosh}(\lambda_{2}\Delta)}{\lambda_{2}}-\dfrac{\textrm{cosh}(\lambda_{3}\Delta)}{\lambda_{3}}\Bigg]^{2}\,.\label{dB-J-W}  
\end{align}
Here $\Lambda=e^{\rho_0}\to \infty$ is the UV cutoff.
The eigenvalues themselves can be expressed compactly in terms of ${\cal L}$ and $C$
\begin{align}
\lambda_{1}&=4\sqrt{\dfrac{2\pi\mathcal{L}}{k}}\sqrt{1-\dfrac{3}{4C}}\label{eigen1}\\
\lambda_{2}&=2\sqrt{\dfrac{2\pi\mathcal{L}}{k}}\Bigg(\sqrt{1-\dfrac{3}{4C}}-\dfrac{3}{2\sqrt{C}}\Bigg)\label{eigen2}\\
\lambda_{3}&=2\sqrt{\dfrac{2\pi\mathcal{L}}{k}}\Bigg(\sqrt{1-\dfrac{3}{4C}}+\dfrac{3}{2\sqrt{C}}\Bigg)\,.\label{eigen3}
\end{align}

To calculate the order $\mu^2$ correction to the entanglement entropy, we first solve eq.(\ref{mu-BTZ-branch}) and express $C$ as a function of $\mu$ and $\mathcal{L}$,
\begin{align}
C=\dfrac{3}{512\pi\mu^{2}\mathcal{L}}\Bigg[c+256\pi\mu^{2}\mathcal{L}+\sqrt{c^{2}+512\pi c\mu^{2}\mathcal{L}}\Bigg]\,.\label{C-val}
\end{align}
Up to this point, everything is exact. In order to express $C$ as a function of the thermodynamical potentials, we need to use the expansion of $\mathcal{L}$ from the holonomy constraints in powers of $\mu$
\begin{align}
\mathcal{L}(\beta,\mu)=\dfrac{\pi c}{12\beta^{2}}+\dfrac{20\pi^{3}c}{9\beta^{4}}\,\mu^{2}\,+\,{\cal O}(\mu^{4})\label{L-exp}
\end{align}
Plugging (\ref{L-exp}) in (\ref{C-val}), $C$ becomes a function solely of the thermodynamic potentials $(\beta,\mu)$, expanded for infinitesimal $\mu$. This in turn yields an expansion for the eigenvalues $\lambda_{1,2,3}$ in powers of $\mu$. Substituting these expansions into the functional (\ref{dB-J-W}) we find at ${\cal O}(\mu^2)$,
\bea
S_{\textrm{EE}}^{\rm holo}=&&\dfrac{c}{3}\,\textrm{ln}\left[\frac{\beta}{\pi\Lambda^{-1}}{\sinh}\left(\frac{\pi\Delta}{\beta}\right)\right]\,+\,c\,\frac{\mu^2}{\beta^2}
\left[\frac{32\pi^2}{9}\,\left(\tfrac{\pi\Delta}{\beta}\right)\,\coth\left(\tfrac{\pi\Delta}{\beta}\right)\,-
\,\frac{20\pi^2}{9}\right.\\\nonumber\\\nonumber
&&\left.-\,\frac{4\pi^2}{3}\csch^2\left(\tfrac{\pi\Delta}{\beta}\right)\,
\left\{\left(\tfrac{\pi\Delta}{\beta}\coth\left(\tfrac{\pi\Delta}{\beta}\right)\,-\,1\right)^2\,+\,\left(\tfrac{\pi\Delta}{\beta}\right)^2\right\}\right]\,+\,{\cal O}(\mu^4)
%\dfrac{c\pi^{2}\mu^{2}}{18\beta^{4}}\,\textrm{csch}^{4}\Bigg(\dfrac{\pi\Delta}{\beta}\Bigg)\Bigg[S_{0}+S_{1}+S_{2}\Bigg]+O(\mu^{4})\,,\label{mu-inf}
%\end{align}
%where
%\begin{align}
%S_{0}&=-24\pi^{2}\Delta^{2}\textrm{cosh}\Bigg(\dfrac{2\pi%\Delta}{\beta}\Bigg)\label{S0}\\
%S_{1}&=+8\pi\beta\Delta\Bigg[\textrm{sinh}\Bigg(\dfrac{2\pi\Delta}{\beta}\Bigg)+\textrm{sinh}\Bigg(\dfrac{4\pi\Delta}{\beta}\Bigg)\Bigg]\label{S1}\\
%S_{2}&=+\beta^{2}\Bigg[-3+8\textrm{cosh}\Bigg(\dfrac{2\pi%\Delta}{\beta}\Bigg)-5\textrm{cosh}\Bigg(\dfrac{4\pi\Delta}{\beta}\Bigg)\Bigg]\,.\label{S2}
\eea
Remarkably, this formula is identical to the result \eqref{EEmain} we have obtained for the free boson and fermion field theories.
Note that as stated  we have observed the agreement with the `holomorphic'  version of 
the higher spin entanglement entropy proposal of 
\cite{deBoer:2013vca}. It is possible that the redefinitions of the charges and chemical potentials given in 
\cite{Compere:2013nba} which renders  the `canonical' thermal entropy formally equivalent to 
the `holomorphic' thermal entropy \footnote{See  for example the equation
(3.49) of the first reference in \cite{Compere:2013nba}.}  might also transform the `canonical' version 
of the entanglement entropy proposed in \cite{deBoer:2013vca, Ammon:2013hba} to the holomorphic 
one used in \eqref{hol-dB-J}. We have performed this excercise and found that the canonical and holomorphic EE proposals do not agree under the variable change of \cite{Compere:2013nba}\footnote{We thank the anonymous referee of this paper for suggesting this potential check.}. It is important to note that, unlike the holomorphic Wilson line functional \eqref{dB-J-W}, the canonical Wilson line functional is not entirely expressible in terms of eigenvalues of components of the gauge-connection.  It would be interesting to explore this issue further and understand the field theoretic origin of the proposed formula for canonical EE.

%This will ensure that under this re-definitions  of the charges and 
%chemical potentials the entanglement entropy evaluated from the CFT will 
%also agree with that obtained from the `canonical' version of the %holographic 
%higher spin entanglement entropy proposal. 
%It is important to verify if this expectation is indeed true. 

We would like to stress that {\em a priori} we have no reason to expect the results for these three different theories to agree. In particular, the gravity calculation we have just outlined applies to a boundary CFT with ${\cal W}_3$-symmetry and with central charge $c\to\infty$. On the other hand our CFT computations were performed in theories with ${\cal W}_\infty[\lambda]$ symmetry for $\lambda=0,1$ and fixed central charges. It is therefore natural to conjecture that perhaps the order $\mu^2$ correction to the EE (with spin-3 chemical potential) is universal for CFTs with ${\cal W}$-symmetries.

\section{Free fermions, U(1) coset and ${\cal W}_N$-algebras}
\label{cosetting}
In this section we explore another possible avenue offered by the free fermion theory to perform a calculation of the ${\cal O}(\mu^2)$ correction to the EE within a free CFT realization of ${\cal W}$-symmetry. Our starting point is the observation made in \cite{Gaberdiel:2013jpa}
that the U(1) current in the ${\cal W}_{1+\infty}$ algebra can be eliminated by a process of cosetting by this current. The procedure is to first construct currents in the CFT with $N$ free fermions which have vanishing OPEs with the 
U(1) current $J$. In the process the central charge of the CFT is reduced to $N-1$.

We first point out that the procedure of cosetting by $J$ has one other very interesting consequence which, to our knowledge, appears not to have been noticed previously. In short, we empirically observe that the U(1)-cosetted algebra for $N$ free fermions appears to truncate to a {\em finite} number of currents and yields a ${\cal W}_N$-algebra.

\subsection{U(1) coset generators}
The generators of the U(1) coset of the ${\cal W}_{1+\infty}[\lambda=0]$ algebra were constructed recursively in \cite{Gaberdiel:2013jpa} to have vanishing OPEs with $J$ and so that higher spin currents are primaries with respect to the new stress tensor. We list the new currents up to spin $s=4$, namely $(\widetilde T, \widetilde W, \widetilde U)$ in terms of the old ones $(T, W, U)$:
\bea
&&\widetilde T\,=\, T\,-\frac{1}{2N}:JJ:\,,\label{cosetgen}\\\nonumber
&&\widetilde W\,=\, W \,-\,\frac{2}{N}:JT:\,+\,
\frac{2}{3N^2}:JJJ:\,,\\\nonumber
&& \widetilde U\,=\, U\,+\,\frac{1}{10N}\,:J\partial^2J:\,-\,\frac{3}{20N}:\partial J\partial J:\,-\,\frac{3}{N}:JW:\,+\,\frac{3}{N^2}:JJT:\,\\\nonumber
&&\qquad-\,\frac{3}{4N^3}:JJJJ: \,-\,\tfrac{21-15/N}{5N+17}\,\left(
:\widetilde T\widetilde T: -\frac{3}{10}\partial^2\widetilde T\right)\,.
\eea
Using the OPE $J(z) J(w)\sim N/(z-w)^2$, it is readily seen from the  $\widetilde T\widetilde T$ OPE that the central charge of the cosetted theory is $c=N-1$. With this definition, the spin-3 and spin-4 currents are primaries with respect to the coset stress tensor.

\subsection{Bosonized formulation and truncation}
The point we want to make is most easily visible in the bosonized picture. We take the normalized currents, after bosonization, to be \cite{Pope:1991ig}:
\bea
&&J\,=\,i\sum_{a=1}^N\partial\varphi_a\,,\quad\quad
T\,=\,-\frac{1}{2}\sum_{a=1}^N (\partial\varphi_a)^2\,\quad\quad
W\,=\,-\frac{i}{3}\sum_{a=1}^N(\partial\varphi_a)^3\,\\\nonumber
&&U\,=\,\sum_{a=1}^N \left[\frac{1}{4}(\partial\varphi_a)^4 - \frac{3}{20}(\partial^2\varphi_a)^2+\frac{1}{10}\partial\varphi_a\partial^3\varphi_a\right]\,.
\eea
For the case of a single free fermion, $N=1$, the cosetting procedure 
\eqref{cosetgen} yields a vanishing stress tensor $\widetilde T =0$ and $\widetilde W=0$ so the algebra is trivially truncated since the theory has no degrees of freedom. When $N=2$, the coset stress tensor is that of a single free boson:
\be
N=2:\qquad\widetilde T\,=\,-\tfrac{1}{2}:(\partial\Phi)^2:\qquad
\Phi\,=\,\frac{\varphi_1-\varphi_2}{\sqrt 2}\,,
\ee 
 and a vanishing spin-three current $\widetilde W=0$. Since 
 $J=i\partial(\varphi_1+\varphi_2)$, it is guaranteed to have a vanishing OPE with the improved stress tensor $\widetilde T$.
 
For $N=3$, we find that $(\widetilde T, \widetilde W)$ are both non-zero and 
$\widetilde U\,\sim\, :\widetilde T\widetilde T:$ so that the spin-4 current is not an 
independent current. Indeed, upon careful evaluation we find that the coset currents $(\widetilde T, \widetilde W)$ precisely have the OPEs of the ${\cal W}_3$ algebra with $c=2$. The truncation to the ${\cal W}_3$ algebra (see appendix \ref{walgebra} for ${\cal W}_3$ OPEs) is easy to see when we define a new basis of fields, and express $(\widetilde T, \widetilde W)$ in this new basis.

We choose a new field basis $(\Phi_1,\Phi_2, \Phi_3)$ related to the old one by an orthogonal transformation
\bea
\begin{pmatrix}
\Phi_1\\
\Phi_2\\
\Phi_3 
\end{pmatrix} = \begin{pmatrix}
\tfrac{1}{\sqrt 3} &\tfrac{1}{\sqrt 3} & \tfrac{1}{\sqrt 3}\\
\tfrac{1}{\sqrt 2} &-\tfrac{1}{\sqrt 2} &0\\
\tfrac{1}{\sqrt 6} &\tfrac{1}{\sqrt 6} &-\tfrac{1}{\sqrt 6}  
\end{pmatrix}\begin{pmatrix}
\varphi_1\\
\varphi_2\\
\varphi_3 
\end{pmatrix} \,.
\eea
In terms of the new fields, the coset currents have the property that they are independent of $\partial\Phi_1$, the `centre of mass mode' which is also the $U(1)$ current $J$. Therefore their OPEs of the coset currents with $J$ vanish trivially.
 Explicitly, the coset stress tensor and $W$-current (normalized as before) are,
\def\cW{\mathcal{W}}
\def\cln{\boldsymbol{\colon}\hspace{-.13cm}}
\def\rn{\hspace{-.08cm}\boldsymbol{\colon}\hspace{-.05cm}}
\bea \label{ff-l}
&&\widetilde T(z) \,=\, -\frac{1}{2} \cln \left( (\pd \Phi_2)^2 (z)\,+\,(\pd \Phi_3)^2 (z)   \right)  \rn  \\\nonumber\\
&&\widetilde W(z) \,=\,- \frac{1}{2\pi}\sqrt{\frac{5}{2}} \cln \left( (\pd \Phi_2)^2\partial\Phi_3(z)\,-\, \frac{1}{3}(\pd \Phi_2)^3 (z) \right) \rn \label{ff-w}
\eea
It can be checked that the OPEs of the above free field realizations are indeed those of the $\mathcal{W}_3$ algebra with $c=2$. This realization of the ${\cal W}_3$ algebra is actually identical to the well known Miura transformation \cite{Fateev:1987vh, Fateev:1987zh, Pope:1991ig}. It is extremely interesting that the U(1) coset of the ${\cal W}_{1+\infty}$ algebra, as proposed in \cite{Gaberdiel:2013jpa} leads to the free field realization of the ${\cal W}_3$-algebra provided by the Miura transform. We have also been able to verify that for general $N$, the form of the spin-3 current in the 
U(1) coset matches the one obtained by Miura transformation.
 
\subsection{Entanglement entropy for ${\cal W}_3$ CFT}
To compute the EE in the above free field realization of 
${\cal W}_3$, we first recall that the latter  was obtained as a U(1) coset of the free fermion theory. The coset currents are not simply bilinears in the fermions, as they were before the procedure. The stress tensor $\widetilde T$ contains a four-fermi term and so the new theory is actually interacting.
The relation between the original free fermions and the new free bosons $\Phi_{1,2}$ is not simple. Instead, we may think of the free bosons $\Phi_{1,2}$ as being related to a new set of free fermions,
\be
\Psi_{1,2}\,=\,:e^{i\Phi_{1,2}}:\,\,
\ee
We can then try to compute the R\'enyi entropy of this system using twist operators appropriate for these new free fermions,
\bea
&&\sigma_{k,n}(y_1,\bar y_1)\,=\,:e^{i\frac{k}{n}(\Phi_1(y_1)+\Phi_2(y_1)-\bar\Phi_1(\bar y_1)-\bar\Phi_2(\bar y_1))}:\\\nonumber\\\nonumber
&&\bar\sigma_{k,n}(y_2,\bar y_2)\,=\,
:e^{-i\frac{k}{n}
(\Phi_1(y_2)+\Phi_2(y_2)-\bar\Phi_1(\bar y_2)-\bar\Phi_2(\bar y_2))}:
\eea
with $k= -\tfrac{1}{2}(n-1),\ldots,\tfrac{1}{2}(n-1)$. We stress that at this stage it is not clear that these are the correct twist fields after the coseting procedure. Our goal is to  see if the basic structure of the entanglement entropy calculation we saw for ${\cal W}_\infty[\lambda=0]$ and ${\cal W}_{\infty}[\lambda=1]$  can be reproduced. 

The subsequent analysis follows the discussion in section \ref{EEREff}. The most important point is the structure of the operator product of two $W$-currents. After performing all the requisite Wick contractions, we find the order $\mu^2$ correction has the form
\bea
S_{\rm EE}^{(2)}\,=\,\frac{5}{6\pi^2}\int d^2z_1\int d^2z_2
\left(\frac{1}{2}H^4(z_1-z_2)G(z_1)G(z_2)-\frac{1}{20}
H^2(z_1-z_2)G^2(z_1)G^2(z_2)\right)\,\nonumber
\eea
where $S_{\rm EE}\,=\,S_{\rm EE}^{(0)}\,+\,\mu^2\,S_{\rm EE}
^{(2)}\,+\ldots$.
Surprisingly, this expression has the same form that we have encountered in the previous two examples. However the overall normalization is off by a factor of $\tfrac{1}{2}$, which means that in the high temperature limit, it yields only one half of the ${\cal O}(\mu^2)$ correction to the thermal entropy. We therefore conclude that our proposal for the twist fields in the coset theory is not quite correct because the latter is effectively an interacting theory after the 
cosetting\footnote{We believe that one possible reason why the naive approach for computing RE/EE does not work is that, after cosetting, the spin-three currents are not bilinears in the fermions and thus the standard trick of Fourier transforming in replica space does not disentangle the field modes with different $k$.}. In addition, we have been unable to generalize this calculation in a consistent fashion for $N > 3$, so as to have the correct high temperature behaviour for the entanglement entropy. It is therefore likely that the twist fields we have proposed are not the correct ones for this system. It would be interesting to find a consistent way to implement the R\'enyi entropy calculation within the free field realization of ${\cal W}_N$  CFTs.

\section{Discussion}
\label{summary}

The goal of this paper was to compute  the single interval, finite temperature 
entanglement entropy in tractable examples of CFTs in two dimensions 
with ${\cal W}$-algebra symmetries, deformed by a spin-three chemical potential $\mu$. 
The primary motivation behind this was to obtain results that could be compared with proposals 
for the same in holographically dual higher spin theories of gravity.  In particular, we considered the free boson and free fermion theories which have ${\cal W}_\infty[\lambda]$  symmetry with $\lambda=1$ and $\lambda=0$ respectively, and found that the order $\mu^2$ correction to the entanglement entropy is unexpectedly identical for the two theories. Furthermore we found that the proposed Wilson line prescription for holographic EE in higher spin theories, applied to ${\rm SL(3)}\times {\rm SL(3)}$ Chern-Simons theory, also yields the same result at ${\cal O}(\mu^2)$, for a theory with ${\cal W}_3$ symmetry and large central charge. This suggests that the specific correction we have computed may be universal.

The reason for this common result across three separate examples remains unclear to us. In particular, despite the matching of EE at order $\mu^2$, we have seen that the R\'enyi entropies for the two field theories considered in this paper are not the same at this order. Thus we  cannot  automatically ascribe the agreement to universality in correlators of spin-three currents with twist fields. In the absence of an immediate explanation, it may be fruitful to look for other theories where a similar perturbative computation (in $\mu$) could be carried out within the CFT. The 
$\mathcal{W}_N$ minimal models or WZW coset CFTs considered in 
the higher-spin/minimal-model duality of \cite{gg} may offer a tractable route, provided the twist-operators (or their OPEs) in these interacting theories and their correlation 
functions with the higher spin currents can be determined. 
Another class of CFT's worth examining are the 3-state Potts model and $Z_N$ lattice 
models both of which admit spin three currents. 
It would be equally interesting to extract explicit results (perturbatively in $\mu$) using the holographic EE proposal of \cite{deBoer:2013vca, Ammon:2013hba} for higher-spin black holes in hs$[\lambda]$ theory for some $\lambda$ (other than $\lambda=-3$).

Another test of the putative universality of the order $\mu^2$ correction to EE would be to consider a simple generalization of the calculations in this paper to the non-static case with $\mu\neq-\bar\mu$. This generalization is easily implemented in the free fermion and free boson CFTs as all the necessary ingredients for the calculation are already contained in this paper. It should also be possible to apply the holographic formula to obtain the result for the non-static spin-three black hole solution (perturbatively in the chemical potentials), and compare with CFT results.

An open end in our work is the analysis of the EE with higher spin chemical potential within free field realizations of 
$\mathcal{W}_N$ CFTs for fixed $N$. We have observed empirically that upon cosetting the theory of $N$ free fermions by the U(1) current, the ${\cal W}_{1+\infty}$ algebra is truncated to ${\cal W}_N$. In the bosonized language the resulting free field realization appears to be identical to the Miura transform. We were not able to obtain the correct entanglement entropy for these theories because identification of the correct twist field is unclear, and this presents an interesting avenue for future work. 

All through this paper we have focussed attention only on the holomorphic formalism. As remarked earlier, it has been pointed out that in the so-called canonical approach to thermodynamics \cite{Henneaux:2013dra}, the higher spin chemical potential should only couple to the temporal component of the higher spin current, and included as a deformation of the Hamiltonian. The corresponding Lagrangian deformation deduced from this Hamiltonian will not match with the holomorphic deformations studied in this paper. 
 It should be possible to implement this prescription directly within the free boson or free fermion CFT and compute the corrections to thermal entropy and EE perturbatively within the canonical formalism. 

Finally, it would be fascinating to learn whether a first principles understanding of the  holographic proposal for the EE in higher spin theories can be achieved via bulk solutions of higher spin gravity that directly compute the higher spin 
R\'enyi entropies. Specifically, this would require a higher spin generalization of the work of \cite{faulkner} wherein handlebody solutions in bulk gravity  could be found and their actions computed to eventually yield holographic EE formulae.

\acknowledgments
We would like to thank Mathias Gaberdiel, Rajesh Gopakumar, Tim Hollowood, Gautam Mandal and Aninda Sinha for enjoyable discussions and comments. 
The work of JRD is partially supported by the Ramanujan
fellowship DST-SR/S2/RJN-59/2009.
SPK would like thank all members of CHEP, Indian Institute of Science, Bangalore, for their warm hospitality, stimulating discussions and support.

\appendix
\section*{Appendix}
\section{${\cal W}$-algebra OPEs} 
\label{walgebra}
We list OPEs involving the stress tensor $T$ and the spin-three current $W$, which are universal for any ${\cal W}$-algebra. These particular OPEs (importantly, the $WW$ OPE) are independent of $\lambda$ for ${\cal W_\infty[\lambda]}$ (see e.g. appendix D of 
\cite{Gaberdiel:2012yb})
\bea
&&T(z)T(w)\sim \frac{c/2}{(z-w)^4}+\frac{2 T(w)}{(z-w)^2}+\frac{T'(w)}{(z-w)}\,,\nonumber\\\nonumber
&& T(z)W(w)\sim\frac{3}{(z-w)^2}W(w)+\frac{1}{z-w}W'(w)\,,\\\nonumber
&&W(z)W(w)\sim\frac{2c/3}{(z-w)^6}+\frac{4T(w)}{(z-w)^4}+\frac{2T'(w)}{(z-w)^3}+\frac{4U(w)+\tfrac{3}{5}T''(w)}{(z-w)^2}+\\
&&\qquad+\frac{2U'(w)+
\tfrac{2}{15}T'''(w)}{(z-w)}\,.\label{opew3}
\eea
Here $U$ is the spin-4 current. For the ${\cal W}_3$ algebra, $U$ is replaced by a composite operator since there is no spin-4 current:
\be
U\to\frac{16}{22+5c}\left(:TT: -\tfrac{3}{10}\,T''\right)\,.
\ee

\section{Integrals on the cylinder}
\label{integrals}
\subsection{${\cal O}(\mu^2)$ thermal correction}
\paragraph{{Direct evaluation:}} 
The integral relevant for the order $\mu^2$ correction to the thermal partition function is,
\bea
&&I(z\equiv \sigma+i\tau)\,=\,\int  \frac{d\sigma}{\sinh^6(\sigma + i\tau\,-\,a)}\,=\\\nonumber\\\nonumber
&&\,-\tfrac{1}{15}\coth(\sigma+i\tau-a)\left(
8\,-\,4\,\csch^{2}(\sigma+i\tau-a)\,+\,3\,\csch^{4}(\sigma+i\tau-a)\right)\,.
\label{intthermal}
\eea
With $0\leq\tau <2\pi$, we need $(I(\infty + i\tau) -I(-\infty+i\tau) )= -16/15$.

\paragraph{{Method of singularities:}}
As summarized in section \ref{holpert}, it is possible to better understand the origin of the result of the integration above by examining the singularities of the integrand. This follows from the fact that we can reconstruct the singly periodic function (on the cylinder) from its behaviour near poles. 
The present example provides a good illustration. Near $z=0$, the function $\csch^6(z)$ has the expansion
\be
\csch^6 z\,=\,\frac{1}{z^6}\,-\,\frac{1}{z^4}\,+\,\frac{8}{15 z^2}\,+{\rm regular}
\ee
Performing the indefinite integral term by term yields:
\be
\int\csch^6 z\,=\,-\frac{1}{5\,z^5}\,+\,\frac{1}{3\,z^3}\,-\,\frac{8}{15 z}\,+{\rm regular}
\label{expint}
\ee
Noting that the integrand has periodcity $i\pi$, we sum over the periodic images of the singular terms above (e.g. using identities involving polygamma functions),
\bea
&&\sum_{n=-\infty}^\infty(z-n\pi i)^{-5}\,=\,\coth z \left(\tfrac{1}{3}\csch^2z+\csch^4z\right)\\\nonumber
&&\sum_{n=-\infty}^\infty(z-n\pi i)^{-3}\,=\,\coth z\,\,\csch^2z
\\\nonumber
&&\sum_{n=1}^\infty\left((z-n\pi i)^{-1}+(z+n\pi i)^{-1}\right)+\frac{1}{z}=\coth z\,.
\eea
Combining these along with the coefficients from 
\eqref{expint}, we deduce
\be
\int\csch^6 z\,=\,\coth z\left(-\tfrac{8}{15}+\tfrac{4}{15}\csch^2z-\tfrac{1}{5}\csch^4 z\right)\,
\ee
which is precisely the result \eqref{intthermal}. Importantly, it is only the term proportional to $\coth z$ originating from the double pole in the integrand at $z=0$, which contributes to the definite integral over the cylinder; all higher pole contributions vanish exponentially at infinity.

Similarly, one can also show that if the integrand (assuming a periodicity with multiples of $i\pi$) has simple poles at say $z=z_1,z_2 ,\ldots z_M$ in the fundamental strip then,
\be
\sum_{i=1}^M\frac{a_i}{z-z_i}\,\to\, \sum a_i\coth(z-z_i)\,.
\ee
Upon integrating, this yields 
\be
\sum_{i=1}^M {a_i}\log\sinh(z-z_i)\,\label{logsinh}
\ee
which diverges linearly for large $z$ unless $\sum_i a=0$. Given this, the result of integration from $-\infty$ to $\infty$ yields the following contribution from simple poles in the integrand.
\be
-2\sum_{i=1}^M a_i z_i\,,\qquad\sum_i a_i=0\,.
\ee
Similarly, for double poles at $z=w_1, w_2, \ldots w_P$ in the integrand
\be
\sim \sum_{i=1}^P\frac{b_i}{(z-w_i)^2},
\ee
after integration from $-\infty$ to $\infty$, these yield
\be
-2\sum_{i=1}^P b_i\,.
\ee 
In summary, when integrating periodic functions (on the cylinder), we only need to know coefficients of the simple and double poles and the locations of the former to write down the answer. This method will stand as a very useful check for the much more complicated nested integrals we study below.

\subsection{${\cal O}(\mu^2)$ correction to EE and ${\cal I}_1(\Delta)$} 
For the entanglement entropy we need to perform the double-integrals in eqs.\eqref{I1} and\eqref{I2}. An economical way of proceeding is via the variable change,
\be
z_\pm=\tfrac{1}{2}(z_1\pm z_2)\,,
\ee
so that
\bea
&&{\cal I}_1(\Delta)\,=\,\frac{\pi^6}{\beta^6}\int\int_{{\mathbb R}\times S^1_\beta} d^2z_+ d^2z_-\frac{2\times4\sinh^4\left(\tfrac{\pi\Delta}{\beta}\right)}{\sinh^4{\left(\tfrac{2\pi z_-}{\beta}\right)}\left(\cosh\left(\tfrac{2\pi(\Delta-z_+)}{\beta}\right)\,-\,\cosh\left(\tfrac{2\pi z_-}{\beta}\right)\right)}\nonumber\\\\\nonumber
&&\times\frac{1}{\left(\cosh\left(\tfrac{2\pi z_+}{\beta}\right)\,-\,\cosh\left(\tfrac{2\pi z_-}{\beta}\right)\right)}\,.
\eea
One must be careful in accounting for the Jacobian associated to the variable change to $z_+$ and $z_-$. This results in an extra factor of 2 (indicated in the numerator) whilst keeping the ranges of $\tau_\pm$ restricted to $[0,\beta]$.

\paragraph{{Integration over $z_+$}:} The indefinite integration over $z_+$ is relatively easy. 
Taking $z_+\equiv \sigma_+ + i \tau_+$ and defining the indefinite integral (over $z_+$ or $\sigma_+$) as ${\cal F}(z_+)$, we find
\bea
&&{\cal F}(z_+)=\int \frac{d\sigma_+}{\left(\cosh\left(\tfrac{2\pi z_+}{\beta}\right)\,-\,\cosh\left(\tfrac{2\pi z_-}{\beta}\right)\right)\left(\cosh\left(\tfrac{2\pi(\Delta-z_+)}{\beta}\right)\,-\,\cosh\left(\tfrac{2\pi z_-}{\beta}\right)\right)}\nonumber\\\\\nonumber
&&=\,-\frac{\beta}{4\pi}{\csch\left(\tfrac{\pi\Delta}{\beta}\right)\csch\left(\tfrac{2 \pi z_-}{\beta}\right)}\left(\log\left[\frac{\sinh (\pi(z_-+z_+)/\beta)}{\sinh(\pi(\Delta+z_- - z_+)/\beta)}\right]\times\right.\\\nonumber\\\nonumber
&&\left.\times\csch\left(\tfrac{\pi(2z_-+\Delta)}{\beta}\right)+\log\left[\frac{\sinh (\pi(z_+-z_-)/\beta)}{\sinh(\pi(\Delta-z_- - z_+)/\beta)}\right]\csch\left(\tfrac{\pi(2z_--\Delta)}{\beta}\right)\right)
\eea

\paragraph{{Check from simple and double poles:}} We can verify the above result by checking the coefficients of the simple and double poles of the integrand in ${\cal F}(z_+)$ (viewed as a function of $z_+$ alone). By inspection, the integrand has 4 simple poles at $(z_+) = \pm (z_-)$ and $z_+ = \pm (z_-) +\Delta$, and no double poles. The residues at the simple poles are easily obtained:
\bea
&&{\rm At}\,\,z_+ = \pm (z_-):\qquad -\frac{\beta}{4\pi}{\csch\left(\tfrac{\pi\Delta}{\beta}\right)\csch\left(\tfrac{2 \pi z_-}{\beta}\right)}\csch\left(\tfrac{\pi(2z_-\,\mp\,\Delta)}{\beta}\right)\nonumber\\\nonumber\\\nonumber
&&{\rm At}\,\,z_+ = \Delta\pm (z_-):\qquad \frac{\beta}{4\pi}{\csch\left(\tfrac{\pi\Delta}{\beta}\right)\csch\left(\tfrac{2 \pi z_-}{\beta}\right)}\csch\left(\tfrac{\pi(2z_-\,\pm\,\Delta)}{\beta}\right)
\eea
When coupled with the prescription around eq.\eqref{logsinh}, we obtain precisely the result written above.

\paragraph{{The second integration:}}After evaluating the limits ${\cal F}(\infty+i\tau_+)-{\cal F}(-\infty+i\tau_+)$, the $\tau_+$ integration becomes trivial, giving an overall factor of $\beta$, and we find 
\bea
&&{\cal I}_1(\Delta)\,=\label{intermediate}\\\nonumber
&&\frac{16\pi^6}{\beta^5}\sinh\left(\tfrac{\pi\Delta}{\beta}\right)
\int_{{\mathbb R}\times S^1_\beta} d^2z\frac{z\,\sinh\left(\tfrac{\pi\Delta}{\beta}\right)\cosh\left(\tfrac{2\pi z}{\beta}\right)
-\frac{\Delta}{2}\cosh\left(\tfrac{\pi\Delta}{\beta}\right)\sinh\left(\tfrac{2\pi z}{\beta}\right)}{\sinh^5\left(\tfrac{2\pi z}{\beta}\right)\sinh\left(\tfrac{\pi (2z+\Delta)}{\beta}\right)
\sinh\left(\tfrac{\pi (2z-\Delta)}{\beta}\right)}\,.
\eea
This final integral can be determined by, once again, performing the $\sigma$ integral ($z\,=\,\sigma+i\tau$) first. We have used Mathematica package to perform the indefinite integral (with respect to $\sigma$) and the resulting expression can be simplified after some effort and the limits evaluated to yield eq.\eqref{I1final}. We have also checked this, as explained in detail below, using the shortcut method of examining the singularities of the integrand.

To begin with, we notice that the integrand naturally splits in two pieces: one is manifestly periodic and the other is not (because of the linear dependence on $z)$:
\be
{\cal I}_1\,=\, \int_{{\mathbb R}\times S^1_\beta} d^2z\,\left(z\, g_1(z)\, +\, g_2(z)\right).
\ee
We can easily evaluate the second term by analyzing the singularities of $g_2(z)$. We first anticipate the fact that integral over $\tau$ (the Euclidean time) will only yield a factor of $\beta$. One way to understand this is that the singularities of the integrand lie on the real axis and therefore shifting the integration contour (along the $\sigma$-coordinate) up and down in the $\tau$-direction (within the fundamental strip) does not change the result as the contour does not cross singularities in the process.

In addition we rescale the non-compact coordinate $\sigma = \beta x/2\pi$ so that
\be
\int_{{\mathbb R}\times S^1_\beta} d^2z\, g_2(z)\,
=\,\frac{\beta^2}{2\pi}\int_{-\infty}^\infty dx\, \,g_2\left(\tfrac{\beta x}{2\pi}\right)\,.
\ee
The integrand has simple poles at $x_\pm\,=\,\pm\pi\Delta/\beta$ with residues (coefficients of the simple pole term)
\be
\boxed{a_{\pm}\,=\,\mp \frac{2\pi^5\Delta}{\beta^3}\csch^4
\left(\tfrac{\pi \Delta}{\beta}\right)\,},
\ee
and a double pole at $x=0$, with coefficient
\be
\boxed{b\,=\,\frac{4\pi^5}{3\beta^3}
\Delta\coth\left(\tfrac{\pi\Delta}{\beta}\right)\left(
3\coth^2\left(\tfrac{\pi\Delta}{\beta}\right)-5\right)\,}.
\ee
With our shortcut method outlined above, we may simply write down the result as
\bea
&&\boxed{\int_{{\mathbb R}\times S^1_\beta} d^2z\, g_2(z)\,=\,-2\,(a_+x_+\,+\,a_-x_-) \,-\, 2 b\,}\\\nonumber
&&\qquad\qquad=\,\frac{2\pi ^5 \Delta}{3 \beta ^4} {\csch}\left(\tfrac{\pi  \Delta }{\beta }\right)^4 \left(12 \pi  \Delta -8 \beta\,  {\sinh}\left(\tfrac{2 \pi  \Delta }{\beta }\right)\,+\,\beta\,{\sinh}\left(\tfrac{4 \pi  \Delta }{\beta }\right)\right)
\eea
\paragraph{{Non-periodic integrand}:} We now turn to the second part of the integral where the integrand is not periodic. This can be handled via integration by parts:
\bea
&&\int_{{\mathbb R}\times S^1_\beta}d^2z\, z\,g_1(z)
\to\beta\int_{-\infty}^\infty dx\, x\, g_2(x)\,
\\\nonumber
&&\qquad\qquad=\,\beta\int_{-\infty}^\infty dx\,\frac{d}{dx}
\left(x\,\int g_2(x)\right)\,-\beta\int_{-\infty}^\infty dx\,\left(\int g_2(x)\right)
\eea
We apply the same procedure as before, except that now {\em single, double and third order poles } in $g_2(x)$ become relevant. This is because of the integration by parts which introduces an extra indefinite integral. Tracking these 3 different contributions we find,
\bea
&&\beta\int_{-\infty}^\infty dx\, x\, g_2(x)\,=\,
16\beta^2\sinh^2\left(\tfrac{\pi\Delta}{\beta}\right)\frac{\pi^6}{\beta^6}\left[-\tfrac{1}{12\pi^2}\left(\csch^2\left(\tfrac{\pi\Delta}{\beta}\right)-3
\csch^4\left(\tfrac{\pi\Delta}{\beta}\right)\right)\right.\nonumber\\
&&\left. + \tfrac{1}{2}\csch^6\left(\tfrac{\pi\Delta}{\beta}\right)\,x\,\log\left(\frac{\sinh(x-\pi\Delta/\beta)\sinh(x+\pi\Delta/\beta)}{\sinh^2x}\right)\Big|_{-\infty}^{\infty}\right.\\\nonumber
&&\left.-\tfrac{1}{2}\csch^6\left(\tfrac{\pi\Delta}{\beta}\right)\,\int_{-\infty}^\infty\log\left(\frac{\sinh(x-\pi\Delta/\beta)\sinh(x+\pi\Delta/\beta)}{\sinh^2x}\right)\right]
\eea
The second line vanishes upon taking the limits appropriately, while the integral in the third line can be calculated analytically and found to be equal to $2\pi^2\Delta^2/\beta^2$. This is done
by expanding the integrand as a power series in exponentials and then performing the integrals:
\bea
&&\lim_{\Lambda\to\infty}\int_{-\Lambda}^\Lambda dx\,
\log 2\sinh(x-a)\,=\lim_{\Lambda\to\infty}\int_{-\Lambda}^a  dx\,\left[-(x-a)+\log(1-e^{2(x-a)})+i\pi\right]\nonumber\\
&& + \lim_{\Lambda\to\infty}\int_{a}^\Lambda  dx\,\left[(x-a)+\log(1-e^{-2(x-a)} )\right]\\\nonumber
\eea
Expanding the logarithms, we find that the above expression becomes (after taking the cutoff $\Lambda$ to $\infty$ in the finite itegrals),
\bea
&&-\tfrac{1}{2}\left(a^2-\Lambda^2\right)+a(a+\Lambda)+i\pi(\Lambda+a)-\sum_{n=1}^\infty\frac{1}{n^2} +
\tfrac{1}{2}\left(\Lambda^2- a^2\right)-a(\Lambda-a)
\nonumber\\
&& \,=\,\left(\Lambda^2+ a^2\right) +i\pi(\Lambda+a)-\frac{\pi^2}{6}\,.
\eea
Including the contribution of all three terms inside the logarithm, the cutoff dependence we obtain the result $2\pi^2\Delta^2/\beta^2$. Therefore we finally have,
\be
\beta\int_{-\infty}^\infty dx\, x\, g_2(x)\,=\,
-\frac{\pi^6}{\beta^6}\frac{4\beta^2}{3\pi^2}
\left(\beta^2-3\beta^2\csch^2\left(\tfrac{\pi\Delta}{\beta}\right)+3\pi^2\Delta^2\csch^4\left(\tfrac{\pi\Delta}{\beta}\right)\right)\,.\nonumber
\ee 
Putting everything together and after some manipulations we can cast the result in the form,
\bea
&&{\cal I}_1\left(\Delta\right)\,=\, \frac{4\pi^4}{3\beta^2}\,\left(\frac{4\pi\Delta}{\beta}\,\coth\left(\tfrac{\pi\Delta}{\beta}\right)\,-\,1\right)\,+\,
\\\nonumber
&&\qquad\qquad+\,\frac{4\pi^4}{\beta^2}\sinh^{-2}\left(\tfrac{\pi\Delta}{\beta}\right)\,\left\{\left(1-\frac{\pi\Delta}{\beta}\coth\left(\tfrac{\pi\Delta}{\beta}\right)\right)^2\,-\,\left(\tfrac{\pi\Delta}{\beta}\right)^2\right\}
\eea
\subsection{${\cal O}(\mu^2)$ correction to EE and ${\cal I}_2(\Delta)$} 
The second set of integrals is obtained from the double pole terms in the OPEs of the spin-3 currents with twist operators. These are also dealt with by first introducing new variables $z_\pm = (z_1\pm z_2)/2$, so that 
\bea
&&{\cal I}_2\,=\,
\frac{\pi^6}{\beta^6}\int \int d^2z_+ d^2z_-\frac{2 }{\sinh^2 \left(\frac{2 \pi }{\beta } z_-\right) }\frac{ 16\,\text{sinh}^4\left(\frac{\pi }{\beta }\Delta \right)}{\left(\cosh \left(\frac{\pi }{\beta }(2 \Delta - 2 z_+)\right) - \cosh \left(\frac{2\pi }{\beta }z_-\right)\right)^2}\times
\nonumber\\
&&\frac{1}{\left(\cosh \left(\frac{2\pi }{\beta } z_+\right) - \cosh \left(\frac{2\pi }{\beta }z_-\right)\right)^2}\,.
\eea
We do not repeat the exercise of explaining the evaluation of this integral explicitly since the steps involved are somewhat more tedious, but the methods discussed above 
again apply and yield eq.\eqref{I2final} as the final result. We have also evaluated the integrals using Mathematica, simplifying the results and evaluating the limits by hand. The  end result is 
\bea
&&{\cal I}_2\left(\Delta\right)\,=\,\frac{8\pi^4}{\beta^2}\,\left(5\,-\,\frac{4\pi\Delta}{\beta}\,\coth\left(\tfrac{\pi\Delta}{\beta}\right)\right)\,+\,\\\nonumber
&&\qquad\qquad+\,\frac{72\pi^4}{\beta^2}\sinh^{-2}\left(\tfrac{\pi\Delta}{\beta}\right)\,\left\{\left(1-\frac{\pi\Delta}{\beta}\coth\left(\tfrac{\pi\Delta}{\beta}\right)\right)^2\,-\,\frac{1}{9}\left(\tfrac{\pi\Delta}{\beta}\right)^2\right\}\,.
\eea

\end{document}